%% file: paper.tex
\documentclass[aps,showpacs,preprint,12pt]{revtex4}
\usepackage{amssymb,amsmath}

\newif\ifpdf
\ifx\pdfoutput\undefined
\pdffalse 
\else
\pdfoutput=1 
\pdftrue
\fi

\ifpdf
\usepackage[pdftex]{graphicx}
\else
\usepackage{graphicx}
\fi

\input macros	
\renewcommand{\COL}{\EMB{col}}
\allowdisplaybreaks

\begin{document}

\ifpdf
\DeclareGraphicsExtensions{.pdf, .jpg, .tif}
\else
\DeclareGraphicsExtensions{.eps, .jpg}
\fi

\title{Procedures for Converting among Lindblad, Kraus and Matrix Representations of Quantum Dynamical Semigroups%
\footnote[0]{\hspace{-\parindent}Submitted to \textit{J.~Math.~Phys.}; available from \texttt{http://arxiv.org/abs/quant-ph/0201127}.}}
\author{\raisebox{0pt}[0.5in][0pt]{Timothy F. Havel}}
\affiliation{Nuclear Engineering, Massachusetts Institute
of Technology, Cambridge, 02139, USA}

\date{\raisebox{0pt}[0pt][0.4in]{May 18, 2002}}

\begin{abstract} \noindent
Given an quantum dynamical semigroup expressed as an exponential superoperator acting on a space of $N$-dimensional density operators, eigenvalue methods are presented by which canonical Kraus and Lindblad operator sum representations can be computed. These methods provide a mathematical basis on which to develop novel algorithms for quantum process tomography, the statistical estimation of superoperators and their generators, from a wide variety of experimental data. Theoretical arguments and numerical simulations are presented which imply that these algorithms will be quite robust in the presence of random errors in the data.
\vspace{0.25in}
\end{abstract}


\maketitle

\section{Introduction} \label{sec:intro}
The statistical estimation of superoperators from experimental data is variously known as ``quantum channel identification'' \cite{Fujiwara:01} or ``quantum process tomography'' (QPT) \cite{NielsenChuang:01}. While this task is important throughout experimental quantum physics, it is an essential component of on-going efforts to develop devices capable of reliable quantum information processing and transmission. At the same time, it is only through these efforts that it is now becoming possible to observe and control quantum systems with the precision needed to collect sufficient data for QPT. At the time of writing, however, very few experimental efforts to systematically determine the complete superoperators of natural or engineered quantum processes have been carried out. An instructive example may be found in \cite{ChiChuLeu:01}, where the QPT procedure detailed in \cite{NielsenChuang:01} was applied to NMR data on the two-qubit molecule chloroform. This was followed by fitting a specific decoherence model to the superoperators thereby obtained at multiple time points, in order to estimate the decoherence rates in the model.

The goal of the present paper is to give a reasonably complete and self-contained account of the mathematics needed for \emph{robust} QPT, assuming for the most part that the quantum dynamics may be aptly modelled as a quantum dynamical semigroup (QDS). A QDS describes the evolution of a general open quantum system under the Born-Markov approximations \cite{AlickiFannes:01,GardZoller:00,Weiss:99}, and as such is sufficient to cover most of the systems currently being used or developed for quantum information processing and transmission. By ``robust'', we mean that the QPT results will not be sensitive to random errors in the data, which is critical since these data are often difficult to obtain and significantly contaminated by noise and other errors. In addition, it is desirable to avoid model fitting and instead to determine the complete superoperator making no prior assumptions about it, although this significantly increases the number of parameters to be estimated.

The robustness of our approach is obtained primarily by using the orthogonal projection of an arbitrary Hermiticity-preserving superoperator or QDS generator onto the convex cone of completely positive superoperators and their generators \cite{AlickiFannes:01,GardZoller:00,Weiss:99}. Of necessity, therefore, this account will rederive much that is already known about quantum dynamical semigroups as well as more general completely positive superoperators, using a consistent notation, fixed operator basis, and a standard set of matrix tools \cite{HornJohnson:91, Lutkepohl:96}. These derivations do not involve qualitative physical arguments (coarse-graining, separation of time scales, etc.), but only the mathematical definitions of the objects involved, and extend much of our earlier work on the ``Hadamard'' representation, which exists for any ``diagonal'' superoperator, to more general completely positive superoperators and QDS generators \cite{HaShViCo:01}.

The main results will be eigenvalue methods by which the projection of an arbitrary Hermiticity-preserving superoperator or QDS generator onto the convex cone of completely positive superoperators can be computed. These projections will be shown to yield certain canonical Kraus and Lindblad representations of completely positive superoperators and QDS generators, respectively, which may be novel and are certainly not well-known. The explicit form of the involution which identifies a Hermiticity-preserving superoperator with a quadratic form (or Hermitian supermatrix), herein denoted by ``$\EMB{Choi}$'', also appears to be new (see Corollary \ref{thm:keycor} ff.). It is not the intention of this paper to give a single fixed recipe for QPT, because any such recipe must depend to some extent on the nature of the data to be analyzed. Nevertheless, a simple example will be given using simulated data plus added random noise, which should make it clear how such recipes can be derived from these results and further demonstrates that such recipes may be expected to be robust.

\section{Background on Quantum Dynamical Semigroups and their Representations} \label{sec:background}
This section provides the essential background on quantum dynamical semigroups needed in the remainder of the paper, and in addition defines the basic mathematical operations and notation to be used throughout the paper. A quantum dynamical semigroup (QDS) \cite{AlickiFannes:01, GardZoller:00, Weiss:99} constitutes a bounded one-parameter family of ``super\-operators'' $\ALG S = \ALG S(\,t\M[0.1];\,\cdot\,)$ acting linearly on a space of self-adjoint ``density'' operators $\rho$, and satisfying $\ALG S(\,t+t'\M[0.1];\,\rho\,) = S(\,t\M[0.1];\,\rho\,) S(\,t'\M[0.1];\,\rho\,)$ for all $\rho$ and $t,\,t'\ge0$. Assuming that $\rho$ acts in turn on a complex Hilbert space of dimension $N < \infty$, a general means of representing a QDS is as a \emph{Kraus operator sum} \cite{Kraus:83}, namely
\begin{equation}
\rho(t) ~\equiv~ \ALG S(\,t\M[0.1];\,\rho\,) ~=~
\sum_{m=0}^{M}\, S_{m\DAB{1.3ex}}(t)\,\rho\,S_m^{\,\dag}(t) ~,
\end{equation}
where one may take $M < N^2$, the $S_m$ act on the same Hilbert space as $\rho = \rho(0)$, and the dagger ($\dag$) denotes the adjoint. This ensures not merely that $\ALG S$ preserves the positive semidefiniteness of the density operator $\rho$, but moreover that it is  \emph{completely positive}, meaning that the trace over any other quantum system on which $\ALG S$ acts trivially is again a positive semidefinite operator, as expected for any physically realizable process [\emph{loc.~cit.}].

On identifying the Kraus operators $S_m$ with a matrix representation thereof $\MAT S_m$, a well-known result regarding Kronecker matrix (or tensor) products \cite{HornJohnson:91, Lutkepohl:96} implies
\begin{equation}
\COL(\RHO(t)) ~=~ \EMB{\ALG S}(t)\, \COL(\RHO) ~\equiv~
\sum_{m=0}^{M}\, \big(\,\OL{\MAT S}_m(t) \otimes \MAT S_m(t)\big)\,
\COL(\RHO) ~,
\label{eq:kropsum}
\end{equation}
where $\COL(\RHO)$ denotes the result of stacking the columns of the corresponding density matrix $\RHO$ in left-to-right order on top of one another to get a single column vector of dimension $N^2$, the overline denotes the complex conjugate, ``$\otimes$'' the Kronecker product and juxtaposition denotes matrix multiplication. Although this result (which can be proved by straightforward index gymnastics) is often neglected in theoretical treatises on open quantum systems, it is extremely useful for computational purposes. In particular, it converts the two-sided operations in the Kraus operator sum to one-sided matrix operations, thereby providing a matrix representation of the one-parameter semigroup $\ALG S$. It further makes clear that a completely general linear transformation $\ALG T$ of the ``Liouville'' (matrix) space $\FLD C^{N\times N}$ can also be written in operator sum form as
\begin{equation}
\MAT X' ~\equiv~ \ALG T(\MAT X) ~=~
\sum_{m,n=0}^{N^2-1}\, \tau_{mn}\MAT T_{\DAB{1.35ex}m}\, \MAT X\, \MAT T_n^\dag ~,
\label{eq:matmap}
\end{equation}
where $\MAT X \in \FLD C^{N\times N}$, the $\MAT T_m$ are a matrix basis thereof, and $\tau_{mn} \in \FLD C$ are coefficients, since
\begin{equation}
\label{eq:opsum}
\COL(\MAT X') ~=~ \Bigg( \sum_{m,n=0}^{N^2-1}\, \tau_{mn}\, \OL{\MAT T}_n \otimes \MAT T_m \Bigg) \COL(\MAT X) ~\equiv~ \EMB{\ALG T}\,\COL(\MAT X) ~,
\end{equation}
and $[\,\OL{\MAT T}_n \otimes \MAT T_m \mid 0\le m,n\le N^2-1]$ constitutes an induced basis for the space of ``supermatrices'' $\FLD C^{N^2\times N^2}$. Clearly, $\ALG T$ preserves Hermiticity if and only if the matrix of coefficients $[\tau_{mn}]_{m,n=0}^{N^2-1}$ is Hermitian.

The semigroup property $\ALG S(\,t+t'\M[0.1];\,\rho\,) = S(\,t\M[0.1];\,\rho\,) S(\,t'\M[0.1];\,\rho\,)$ is of course not assured by the existence of a Kraus operator sum representation, but it is equivalent to the existence of a constant superoperator $\ALG G \in \FLD C^{N^2\times N^2}$ such that $\ALG S(t) = \mathit{Exp}(-\ALG G\,t)$ for all $t\ge0$, where ``$\mathit{Exp}(-\ALG G\,t) = \ALG I - \ALG G\,t + \HALF \ALG G^2\,t^2 + \cdots$'' is the corresponding exponential superoperator (see Ref.~\cite{NajfeHavel:95a} and citations therein). In general, however, such an exponential will not possess a Kraus operator sum representation, even if the real parts of the eigenvalues of $\ALG G$ are nonnegative (ensuring that the evolution is bounded). General necessary and sufficient conditions for a bounded one-parameter family of superoperators to be a QDS were first derived independently by Lindblad \cite{Lindblad:76} and by Gorini, Kossakowski and Sudarshan \cite{GorKosSud:76}, and require that the derivative $\dot\rho$ can be written in the so-called \emph{Lindblad form},
\begin{equation}
\dot\rho(t) ~=~ \ALG L(\rho) ~\equiv~  \imath\big[ \rho(t), H \big] \,+\, \sum_{m=0}^M\, \Big( L_{\X[1.35]m}\, \rho(t)\, L_m^{\dag} \,-\, \HALF\, L_m^{\dag}\, L_{\X[1.35]m}\, \rho(t) \,-\, \HALF\, \rho(t)\, L_m^{\dag}\, L_{\X[1.35]m} \Big) ~,
\end{equation}
where $M < N^2$ as above, and both the Hamiltonian $H$ and Lindblad operators $L_m$ are time-independent. The superoperator $\ALG L$ itself is called the \emph{Lindbladian}. Translated into matrices, this implies that the decoherent part $\EMB{\ALG G}$ of the Lindbladian can be written as
\begin{equation}
\label{eq:lindgen}
\EMB{\ALG G} ~=~ -\sum_{m=0}^M\, \Big( \OL{\MAT L}_m \otimes \MAT L_m \,-\, \HALF\, \MAT I \otimes \big( {\X[1.85]\MAT L}_m^{\dag}\, \MAT L_{\X[1.35]m} \big) \,-\, \HALF\, \big( \OL{\MAT L}_ m^{\,\dag}\, \OL{\MAT L}_{\X[1.35]m} \big) \otimes \MAT I \Big) ~,
\end{equation}
where $\MAT I$ is the $N\times N$ identity matrix.

\section{A Canonical Kraus Operator Sum Representation} \label{sec:CKOSR}
Although superoperators on Liouville space can be represented with respect to an arbitrary supermatrix basis, as in Eq.~(\ref{eq:matmap}), any Liouville space basis induced by an arbitrary Hilbert space basis can be regarded as the basis of elementary matrices $\MAT E_{ij}$ (with a ``$1$'' in the $ij$-th position and zeros elsewhere), which has the advantage identifying the coefficients and the supermatrix elements. For example, one can write the transpose of an arbitrary $N\times N$ matrix in operator sum form as
\begin{equation}
\MAT X^\top ~=~ \sum_{i,j=0}^{N-1}\, \MAT E_{\X[1.35]ij}\, \MAT X\, \MAT E_{ji}^\dag ~=~ \sum_{i,j=0}^{N-1}\, \MAT E_{ij}\, \MAT X\, \MAT E_{ij} ~,
\end{equation}
or equivalently, as
\begin{equation}
\COL\big(\MAT X^\top\big) ~=~ \Big( \sum_{i,j=0}^{N-1}\, \MAT E_{ji}\otimes\MAT E_{ij} \Big) \COL(\MAT X) ~\equiv~ \EMB{\ALG K}\, \COL(\MAT X) ~.
\label{eq:kaydef}
\end{equation}
The supermatrix $\EMB{\ALG K}$ plays a important role in what follows. It is easily seen to be both symmetric and orthogonal, i.e.~involutory. Using the relation $\MAT E_{ij} = \MAT e_{\X[1.45]\M[0.05]i\,} \MAT e_j^\top$ (where $\MAT e_{\X[1.25]\M[0.05]i\,},\,\MAT e_j$ are the elementary unit column vectors) together with the mixed product formula $(\MAT A \otimes \MAT B) (\MAT C \otimes \MAT D) = \MAT{AC} \otimes \MAT{BD}$ \cite{HornJohnson:91}, we can also show that it has the interesting property of swapping the order of the factors in a Kronecker product, since
\begin{align}
\EMB{\ALG K}\, (\MAT X \otimes \MAT Y)\, \EMB{\ALG K} ~=~ \notag
& \Big( \sum_{i,j=0}^{N-1}\, \MAT E_{ji} \otimes \MAT E_{ij} \Big)\, \big(\MAT X \otimes \MAT Y\big)\, \Big(\sum_{k,\ell=0}^{N-1}\, \MAT E_{\ell k} \otimes \MAT E_{k\ell}\Big) \\ \notag =~
& \sum_{i,j,k,\ell=0}^{N-1}\, \big(\MAT E_{ji}\,\MAT X\,\MAT E_{\ell k}\big) \otimes \big(\MAT E_{ij\,}\MAT Y\,\MAT E_{k\ell}\big) \\ \\[-\baselineskip] \notag =~
& \sum_{i,j,k,\ell=0}^{N-1}\,\big( \MAT e_j (\MAT e_i^{\top} \MAT X\, \MAT e_\ell)\, \MAT e_k^\top \big) \otimes \big( \MAT e_i (\MAT e_j^{\top}\MAT Y\,\MAT e_k)\, \MAT e_\ell^\top \big) \\ \notag =~
& \Big( \sum_{j,k=0}^{N-1}\, Y_{jk}\, \MAT E_{jk} \Big) \otimes \Big(  \sum_{i,\ell=0}^{N-1}\, X_{i\ell}\, \MAT E_{i\ell} \Big) ~=~ \MAT Y \otimes \MAT X ~,
\end{align}
where the matrix elements have been denoted by $X_{i\ell} \equiv \MAT e_i^{\top} \MAT X\, \MAT e_\ell$ and $Y_{jk} \equiv \MAT e_j^{\top}\MAT Y\,\MAT e_k$.

We now use the relation $\COL(\M[0.1]\MAT x \MAT y^\top) ~=~ \MAT y \otimes \MAT x$ for
arbitrary column vectors $\MAT x$, $\MAT y$ to show how the matrix $\EMB{\ALG K}$ also gives us the $\COL$ of a Kronecker product of matrices as a Kronecker product of their respective $\COL$'s.
\begin{LEMMA}
\label{thm:rearrange}
Given any two $N\times N$ matrices $\MAT X$, $\MAT Y$, we have
\begin{equation}
\COL(\MAT X) \otimes \COL(\MAT Y) ~=~ \big(\, \MAT I \otimes \EMB{\ALG K} \otimes \MAT I \,\big)\, \COL\big(\, \MAT X \otimes \MAT Y \,\big) ~,
\end{equation}
where $\EMB{\ALG K}$ is defined as in Eq.~(\ref{eq:kaydef}).
\end{LEMMA}
\PROOF Applying the definitions, we obtain:
\begin{align}
\big(\, \MAT I \otimes \EMB{\ALG K} \otimes \MAT I \,\big)\, \COL\big(\, \MAT X \otimes \MAT Y \,\big) ~=~ \notag
& \sum_{i,j=0}^{N-1}\, \Big( \big( \MAT I \otimes \MAT E_{ji} \big) \otimes \big( \MAT E_{ij} \otimes \MAT I \big) \Big)\, \COL\big(\, \MAT X \otimes \MAT Y \,\big) \\ =~ \notag
& \COL\Bigg( \sum_{i,j=0}^{N-1}\, \big( \MAT E_{ij} \otimes \MAT I \big)\, \big(\, \MAT X \otimes \MAT Y \,\big)\, \big( \MAT I \otimes \MAT E_{ij} \big)\! \Bigg) \\ =~ 
& \COL\Bigg( \sum_{i,j=0}^{N-1}\, \Big( \big( \MAT E_{ij}\, \MAT X \big) \otimes \big( \MAT Y\, \MAT E_{ij} \big) \Big)\! \Bigg) \\ =~ \notag
& \COL\Bigg( \sum_{i,j=0}^{N-1}\, \Big( \big( \MAT e_i\, (\MAT e_j^\top\MAT X) \big) \otimes \big( (\MAT Y\, \MAT e_i)\, \MAT e_j^\top \Big)\! \Bigg) \\ =~ \notag
& \COL\!\left(\! \bigg( \sum_{i=0}^{N-1}\, \big( \MAT e_i \otimes (\MAT Y\, \MAT e_i) \big) \! \bigg) \bigg( \sum_{j=0}^{N-1}\, \big( (\MAT e_j^{\top\,}\MAT X) \otimes \MAT e_j^\top \big)\! \bigg)\! \right) \\ =~ \notag
& \COL\Big(\! \COL\big( \MAT Y \big) \, \COL^{\top\!} \big( \MAT X \big) \! \Big) ~=~ \COL\big( \MAT X \big) \otimes \COL\big( \MAT Y \big) \notag
\end{align}
QED

\begin{COROLLARY}
\label{thm:keycor}
With everything defined as in the Lemma,
\begin{equation}
\COL(\M[0.1] \MAT Y\M[0.1] )\,\COL^{\,\dag}(\M[0.1] \MAT X\M[0.1] ) ~=~ \sum_{i,j=0}^{N-1}\, \big(\M[0.1] \MAT E_{ij} \otimes \MAT I \M[0.1] \big) \big(\M[0.1] \OL{\MAT X} \otimes \MAT Y \big) \big(\M[0.1] \MAT I \otimes \MAT E_{ij} \big) ~. \label{eq:supsupop}
\end{equation}
\end{COROLLARY}
\PROOF Just apply the inverse of the $\COL$ operation to the second and last lines of the proof of the Lemma, and add a complex conjugation to account for our use of ``$\dag$'' instead of ``$\top$''. QED

The ``super-superoperator'' on the right-hand side of Eq.~(\ref{eq:supsupop}) maps any $N^2 \times N^2$ supermatrix $\EMB{\ALG S}$, acting on $N \times N$ matrices $\MAT X$ as $\EMB{\ALG S}\,\COL(\MAT X)$, to a new supermatrix $\EMB{\ALG T} = \sum_{i,j=0}^{N-1} (\MAT E_{ij} \otimes \MAT I)\, \EMB{\ALG S}\, (\MAT I \otimes \MAT E_{ij})$, the elements of which are a permutation of those of $\EMB{\ALG S}$. The Corollary shows that if $\EMB{\ALG S}$ is a sum of Kronecker products, as in Eq.~(\ref{eq:kropsum}), then $\EMB{\ALG T}$ is a sum of the corresponding rank one dyadic products, as in Eq.~(\ref{eq:supsupop}). Thus, while Eq.~(\ref{eq:opsum}) allows us to construct a supermatrix representation from an operator sum, we are now able to give a procedure for going in the other direction.
\begin{PROPOSITION}
\label{thm:toopsum}
Let $\EMB{\ALG S}, \EMB{\ALG T} \in \FLD C^{N^2\times N^2}$ with $\EMB{\ALG T} = \sum_{i,j=0}^{N-1}\, ( \MAT E_{ij} \otimes \MAT I \big) \EMB{\ALG S}\, ( \MAT I \otimes \MAT E_{ij} )$, and let
\begin{equation}
\EMB{\ALG T} ~\equiv~ \EMB{\ALG V}\,\EMB\Omega\,\EMB{\ALG W}^\dag ~=~ \sum_{k=0}^{N^2-1}\, \omega_k\, \EMB v_{\X[1.44]k}\, \EMB w_k^\dag ~,
\end{equation}
be the singular value decomposition of $\EMB{\ALG T}$ (where $\EMB v_k, \EMB w_k$ are the columns of the unitary supermatrices $\EMB{\ALG V}, \EMB{\ALG W}$, respectively, and $\omega_k \ge 0$ are the singular values). Then for any $\MAT X \in \FLD C^{N\times N}$,
\begin{equation}
\EMB{\ALG S}\,\COL(\MAT X) ~=~ \COL\big(\EMB{\ALG T} \vartriangleright \MAT X\big) ~\equiv~ \COL\Bigg( \sum_{k=0}^{N^2-1}\, \omega_k\, \MAT V_k\, \MAT X\, \MAT W_k^\dag \Bigg) ~,
\end{equation}
where $\COL(\MAT V_k) = \EMB v_{k\,}$, $\COL(\MAT W_k) = \EMB w_k$ and the symbol ``$\,\vartriangleright\!$'' should be read as ``applied to''.
\end{PROPOSITION}
\PROOF This follows immediately from our foregoing observations, together with the fact that the super-superoperator is involutory, since
\begin{equation} \begin{split}
& \sum_{k,\ell=0}^{N-1}\, \big( \MAT E_{k\ell} \otimes \MAT I \big) \Bigg( \sum_{i,j=0}^{N-1}\, \big( \MAT E_{ij} \otimes \MAT I \big)\, \EMB{\ALG S}\, \big( \MAT I \otimes \MAT E_{ij} \big) \Bigg) \big( \MAT I \otimes \MAT E_{k\ell} \big) \\ =\M
& \sum_{i,j,k,\ell=0}^{N-1} \delta_{i\ell}\, \delta_{jk}\, \big( \MAT E_{kj} \otimes \MAT I \big)\, \EMB{\ALG S}\, \big( \MAT I \otimes \MAT E_{i\ell} \big) \\ =\M
& \bigg( \sum_{j=0}^{N-1}\, \MAT E_{jj} \otimes \MAT I \bigg)\, \EMB{\ALG S}\, \bigg( \MAT I \otimes \sum_{i=0}^{N-1}\, \MAT E_{ii} \bigg) ~=~ \EMB{\ALG S} ~.
\end{split} \end{equation}
\leftline{QED}\pagebreak[2]\par\noindent
The matrices $\{\MAT V_k\}$ and $\{\MAT W_k\}$ are not generally unitary, but each of these two sets forms a basis for $\FLD C^{N\times N}$, and each is orthonormal with respect to the Hilbert-Schmidt (or Frobenius) inner product $\langle \MAT X,\MAT Y \rangle \equiv \TR(\MAT X^\dag\MAT Y)$. By expanding the right-matrices $\MAT W_k$ as linear combinations of the left $\MAT V_k$, one can rewrite the action of $\EMB{\ALG T}$ on $\MAT X$ in the more symmetric form given in Eq.~(\ref{eq:matmap}). Thus we have obtained a general means of converting a supermatrix $\EMB{\ALG S}$ acting on columnized matrices $\COL(\MAT X)$ to operator sum form. Much of the foregoing can of course be extended to nonsquare matrices $\FLD C^{M\times N}$, but we shall have no need of that here.

In the case that $\EMB{\ALG S}$ is a (super)matrix representation of a quantum dynamical semigroup, the matrix $\EMB{\ALG T}$ derived from it has considerably more structure, as we shall now show.
\begin{PROPOSITION}
\label{thm:smat}
With everything defined as in Proposition \ref{thm:toopsum}, the derived supermatrix $\EMB{\ALG T}$ can be written as
\begin{equation}
\EMB{\ALG T} ~=~ \sum_{i,j=0}^{N-1}\, \COL(\MAT S_{ij}) \, \COL^\top(\MAT E_{ij}) ~=~ \left[ \X[2.2] \ALG S(\MAT E_{ij}) \right]_{i,j=0}^{N-1} ~\equiv~ \begin{bmatrix}
\ALG S(\MAT E_{11}) & \ALG S(\MAT E_{12}) & \hdots \\
\ALG S(\MAT E_{21}) & \ALG S(\MAT E_{22}) & \hdots \\
\vdots & \vdots & \ddots \end{bmatrix} ,
\label{eq:smat}
\end{equation}
where $\ALG S(\MAT E_{ij}) \in \FLD C^{N\times N}$ is defined by $\COL(\ALG S(\MAT E_{ij})) = \EMB{\ALG S}\,\COL(\MAT E_{ij})$, and $\MAT S_{ij}$ is the $ij$-th $N\times N$ block of the supermatrix $\EMB{\ALG S}$.
\end{PROPOSITION}
\vspace{-15pt} \PROOF
The first equality in Eq.~(\ref{eq:smat}) follows immediately from Corollary \ref{thm:keycor} together with the obvious fact that $\EMB{\ALG S} = \sum_{i,j=0\,}^{N-1} \MAT E_{ij} \otimes \MAT S_{ij}\,$. To prove the second, we first note that for any $0 \le k,\ell < N$,
\begin{equation} \begin{split}
\EMB{\ALG S}\, \COL(\MAT E_{k\ell}) ~=~
& \bigg( \sum_{i,j=0}^{N-1}\, \MAT E_{ij} \otimes \MAT S_{ij} \bigg)
\COL(\MAT E_{k\ell}) \\ =~
& \;\COL\bigg( \sum_{i,j=0}^{N-1}\, \MAT S_{ij\,} \MAT E_{k\ell\,} \MAT E_{ji} \bigg)
~=~\; \COL\bigg( \sum_{i=0}^{N-1}\, \MAT S_{i\ell\,} \MAT E_{ki} \bigg) ~.
\end{split} \end{equation}
It follows that
\begin{equation} \begin{split}
\left[ \X[2.2] \ALG S(\MAT E_{k\ell}) \right]_{k,\ell=0}^{N-1} ~=~ &
\sum_{k,\ell=0}^{N-1}\, \MAT E_{k\ell} \otimes \ALG S(\MAT E_{k\ell})
~=~ \sum_{k,\ell=0}^{N-1}\, \MAT E_{k\ell} \otimes
\Bigg( \sum_{i=0}^{N-1}\, \MAT S_{i\ell} \MAT E_{ki} \Bigg)
\\ =\;\; &
\sum_{i,k,\ell=0}^{N-1}\,\bigl( \MAT I \otimes \MAT S_{i\ell} \bigr)
\bigl( \MAT E_{k\ell} \otimes \MAT E_{ki} \bigr) ~.
\end{split} \end{equation}
On the other hand,
\begin{align}
& \EMB{\ALG T} ~=~ \sum_{i,\ell=0}^{N-1}\, \COL(\MAT S_{i\ell})\, \COL^\top\!(\MAT E_{i\ell})  ~=~ \sum_{i,\ell=0}^{N-1}\, \big( \MAT I \otimes \MAT S_{i\ell} \big)\, \COL(\MAT I)\, \COL^\top\!(\MAT E_{i\ell}) \notag \\
=\;\; & \sum_{i,\ell=0}^{N-1}\, \big( \MAT I \otimes \MAT S_{i\ell} \big) \Bigg( \sum_{k\,=\,0}^{N-1}\, \MAT e_k \otimes \MAT e_k \Bigg) \big( \MAT e_\ell^\top \otimes \MAT e_i^\top \big)  ~=\;\; \sum_{i,\ell,k=0}^{N-1}\, \big( \MAT I \otimes \MAT S_{i\ell} \big) \big( \MAT E_{k\ell} \otimes \MAT E_{ki} \big)
\end{align}
QED \pagebreak[2]

In the form $\sum_{k,l\,} \COL(\MAT S_{k\ell}) \COL^\top(\MAT E_{k\ell})$ the derived supermatrix $\EMB{\ALG T}$ appears to have first been studied in connection with superoperators by Jordan and Sudarshan \cite{JordaSudar:61}, whereas the form $\big[ \ALG S(\MAT E_{ij} \big]_{i,j=0}^{N-1}$ was first used to give an intrinsic characterization of completely positive superoperators by Choi \cite{Choi:75}. For this reason we shall henceforth denote it by $\EMB{Choi}(\EMB{\ALG S}) \equiv \sum_{i,j=0\,}^{N-1} (\MAT E_{ij} \otimes \MAT I) \EMB{\ALG S} (\MAT I \otimes \MAT E_{ij})$. The next Lemma will enable us to show that in the cases of interest here, it is a Hermitian matrix.
\begin{LEMMA}
\label{thm:herm}
A superoperator $\ALG S$ commutes with the operation of taking its adjoint, i.e.~$\ALG S(Z^\dag) = \big(\ALG S(Z)\big)^\dag$ for all operators $Z$ in its domain, if and only if it maps self-adjoint operators to self-adjoint operators, and if and only if for any matrix representation $\EMB{\ALG S}$ of $\ALG S$,
\begin{equation}
\label{eq:herm}
\OL{\!\EMB{\ALG S}} ~=~ \EMB{\ALG K}\, \EMB{\ALG S}\, \EMB{\ALG K} ~,
\end{equation}
where the overbar denotes the complex conjugate and $\EMB{\ALG K}$ is defined as in Eq.~(\ref{eq:kaydef}).
\end{LEMMA}
\PROOF Clearly if $\ALG S$ commutes with the adjoint, it maps self-adjoint operators to the same. Now suppose that $\EMB{\ALG S}$ is a matrix representation of $\ALG S$, and let $\MAT X \in \FLD C^{N\times N}$ satisfy $\MAT X = \MAT X^\dag$; then
\begin{equation}
\COL(\MAT X) ~=~ \EMB{\ALG K}\,\COL(\MAT X^\top) ~=~ \EMB{\ALG K}\,\COL(\OL{\MAT X})
\end{equation}
and hence if $\EMB{\ALG S}$ preserves Hermiticity,
\begin{equation} \begin{split}
\EMB{\ALG S}\, \COL(\MAT X) ~\equiv\M
& \COL(\MAT Y) ~=~ \EMB{\ALG K}\, \COL(\,\OL{\!\MAT Y\!}\,) \\ ~\equiv\M
& \EMB{\ALG K}\; \OL{\!\EMB{\ALG S}}\, \COL(\,\OL{\!\MAT X\!}\,) ~=~ \EMB{\ALG K}\; \OL{\!\EMB{\ALG S}}\, \EMB{\ALG K}\, \COL(\MAT X) ~.
\end{split} \end{equation}
Letting $\MAT X$ range over any Hermitian basis of $\FLD C^{N\times N}$ now proves Eq.~(\ref{eq:herm}). And finally, if $\MAT Z \in \FLD C^{N\times N}$ is any (not necessarily Hermitian) matrix and $\EMB{\ALG S}$ satisfies Eq.~(\ref{eq:herm}), we have
\begin{equation} \begin{split}
\COL\big( \ALG S(\MAT Z^\dag) \big) ~\equiv~ \EMB{\ALG S}\, \COL(\MAT Z^\dag) ~=\M
& \EMB{\ALG S}\, \EMB{\ALG K}\, \COL(\,\OL{\!\MAT Z\!}\,) \\ ~=\M
& \EMB{\ALG K}\; \OL{\!\EMB{\ALG S}}\, \COL(\,\OL{\!\MAT Z\!}\,) ~\equiv~ \COL\big( ({\ALG S(\MAT Z))}^\dag \big)
\end{split} \end{equation}
which, since it holds for any representation $\EMB{\ALG S}$ and matrix $\MAT Z$, proves $\ALG S(Z^\dag) = {(\ALG S(Z))}^\dag$. QED

\begin{COROLLARY}
\label{thm:supadj}
If a superoperator $\ALG S$ commutes with the adjoint operation on its domain, then any Choi matrix for it is Hermitian.
\end{COROLLARY}
\PROOF Clearly a Choi matrix $\big[\ALG S(\MAT E_{ij})\big]_{i,j=0}^{N-1}$ is Hermitian if and only if $\ALG S(\MAT E_{ij}) = {(\ALG S(\MAT E_{ji}))}^\dag$ for all $0\le i,j < N$, and if $\EMB{\ALG S}$ is the corresponding matrix representation of $\ALG S$, Lemma \ref{thm:herm} implies
\begin{equation} \begin{split}
\COL\Big( {(\ALG S(\MAT E_{ji}))}^\dag \Big) ~=\M
& \EMB{\ALG K}\, \COL\big(\, \OL{\!\ALG S}(\MAT E_{ji}) \big) ~=~ \EMB{\ALG K}\; \OL{\!\EMB{\ALG S}}\, \COL( \MAT E_{ji} ) \\ =\M
& \big( \EMB{\ALG K}\; \OL{\!\EMB{\ALG S}}\, \EMB{\ALG K} \big)\, \COL( \MAT E_{ij} ) ~=~ \EMB{\ALG S}\, \COL( \MAT E_{ij} ) ~=~ \COL\big( \ALG S(\MAT E_{ij}) \big) ~.
\end{split} \end{equation}
QED

\begin{THEOREM}
\emph{(Choi \cite{Choi:75})}\label{thm:choi}
Let $\ALG S$ be a superoperator which commutes with the adjoint operation on its domain. Then $\ALG S$ is completely positive if and only if the Choi matrix associated with any matrix representation of $\ALG S$ is positive semidefinite.
\end{THEOREM}
\PROOF Let $\EMB{\ALG S}$ be a matrix representation of $\ALG S$ and $\EMB{\ALG T} = \EMB{Choi}( \EMB{\ALG S} )$ be its Choi matrix. This is Hermitian by Corollary \ref{thm:supadj}, and accordingly, we let
\begin{equation}
\EMB{\ALG T} ~=~ \EMB{\ALG U}\, \EMB{\Xi}\, \EMB{\ALG U}^{\,\dag} ~=~ \sum_{n=0}^{N^2-1}\, \xi_n\, \EMB u_{\X[1.35]n}\, \EMB u_n^{\dag}
\end{equation}
be its eigenvector decomposition, where $\EMB{\ALG U}$ is unitary and the eigenvalues $\xi_n$ are real. Then if $\xi_n \ge0$ for $0 \le n < N^2-1$, we let $\MAT T_n$ be the sequence of $N \times N$ matrices such that $\COL(\MAT T_n) = \sqrt{\xi_n}\, \EMB u_{n\,}$. It now follows from Proposition \ref{thm:toopsum} that
\begin{equation}
\EMB{\ALG S}\, \COL(\RHO) ~=~ \Bigg( \sum_{n=0}^{N^2-1}\; \OL{\MAT T}_n \otimes \MAT T_n \Bigg) \COL(\RHO) ~=~ \COL\Bigg( \sum_{n=0}^{N^2-1}\, \MAT T_{\X[1.35]n}\, \RHO\, \MAT T_n^{\,\dag} \Bigg) ~.
\end{equation}
The right-hand side provides a Kraus operator sum representation for $\ALG S$, which by the previously mentioned work of Kraus \cite{Kraus:83} proves that $\ALG S$ is completely positive, as claimed.

Conversely, if $\ALG S$ is completely positive, it may be expressed in Kraus operator sum form as
\begin{equation}
\ALG S(\rho) ~=~ \sum_{m=0}^M\, S_{\X[1.35]m}\, \rho\, S_m^{\,\dag} ~,
\end{equation}
and it follows from Eq.~(\ref{eq:kropsum}) that any matrix representation $\EMB{\ALG S}$ thereof satisfies
\begin{equation}
\EMB{\ALG S} ~=~ \sum_{m=0}^M\; \OL{\MAT S}_m\, \otimes \MAT S_m
\end{equation}
for suitable $\MAT S_m \in \FLD C^{N\times N}$. By Proposition \ref{thm:toopsum}, therefore, corresponding Choi matrix $\EMB{\ALG T}$ is a sum of dyads, that is
\begin{equation}
\EMB{\ALG T} ~=~ \sum_{m=0}^M\, \COL(\,\MAT S_m\,)\, \COL^\dag(\,\MAT S_m\,) ~,
\end{equation}
which is necessarily positive semidefinite. QED

\begin{COROLLARY}
\label{thm:main}
Any Kraus operator sum $\ALG S(\rho) ~=~ \sum_{m=0}^M\, S_{\DAB{1.35ex}m}\, \rho\, S_m^{\,\dag}$ can be written in canonical operator sum form as
\begin{equation}
\ALG S(\rho) ~=~ \sum_{n=0}^{N^2-1}\, T_{n\X[1.4]}\, \rho\, T_n^{\,\dag} ~,
\end{equation}
with $\big\langle T_n, T_{n'} \big\rangle = \TR(T_n^{\,\dag}\,T_{\X[1.35]n'}) = 0$ for all $0 \le n \ne n' < N^2$ and $\|T_n\|^2 = \big\langle T_n, T_n \big\rangle = 0$ for all $n > M$. Subject to this condition, the canonical form is unique up to the overall phase of the operators $T_n$ unless the Hilbert-Schmidt norms satisfy $\|T_n\| = \|T_{n'}\|$ for some $n' \ne n$, in which case it is only unique up to unitary linear combinations of the operators in such degenerate subspaces.
\end{COROLLARY}
\PROOF Implicit in the proof of Theorem \ref{thm:choi}. QED

\bigskip
\section{\hspace*{0.45em}A CANONICAL LINDBLAD REPRESENTATION%
}
We now turn our attention specifically to quantum dynamical semigroups, which (as mentioned in the Introduction) may be assumed to be given in the form of a superoperator exponential $\ALG S = {Exp}( -\ALG F\, t )$. The time-independent generator will usually be of the form $\ALG F = \ALG G + \imath\,\ALG H$ for superoperators $\ALG G$ and $\ALG H$, where $\imath^2 = -1$, $\ALG H(\rho) ~=~ \big[\,\rho,\,H\,\big]$ for the Hamiltonian $H$ of the system in question, and $\ALG G$ is known as the \emph{relaxation superoperator} \cite{ErnBodWok:87}. Although $\ALG G$ may often be self-adjoint, this is not necessarily the case.

An important property of physically meaningful operations on density operators $\rho$, which we have neglected up to now, is that they preserve the trace $\TR(\rho) = 1$. Given an operator sum representation $\ALG S(\rho) = \sum_{m,n=0}^{N^2-1} \tau_{mn}\, T_{\X[1.35]m}\, \rho\, T_n^{\,\dag}$, this is easily seen to be equivalent to $\sum_{m,n=0}^{N^2-1} \tau_{mn}\, T_n^{\,\dag}\,T_{\X[1.35]m} = I$, the identity. We seek an equivalent condition in terms of a given matrix representation $\EMB{\ALG S}$. To this end we expand $\ALG S$ versus the basis of elementary matrices as
\begin{equation}
\EMB{\ALG S} ~\equiv~ \sum_{i,j,k,\ell=0}^{N-1} s^{ij}_{k\ell}\; \big( \MAT E_{\ell j} \otimes \MAT E_{ki} \big) ~=~ \sum_{i,j,k,\ell=0}^{N-1} s^{ij}_{k\ell}\; (\MAT e_\ell \otimes \MAT e_k)\, {(\MAT e_j \otimes \MAT e_i)}^\dag
\end{equation}
where
\begin{equation}
s^{ij}_{k\ell} ~\equiv~ \TR\big( {(\MAT E_{\ell j} \otimes \MAT E_{ki})}^\dag\, \EMB{\ALG S} \big) ~=~ {(\MAT e_\ell \otimes \MAT e_k)}^\dag\, \EMB{\ALG S}\,(\MAT e_j \otimes \MAT e_i) ~,
\end{equation}
so that the corresponding operator sum representation becomes
\begin{equation}
\EMB{\ALG S}\, \COL(\RHO) ~=~ \COL\Bigg( \sum_{i,j,k,\ell=0}^{N-1} s^{ij}_{k\ell}\;
\MAT E_{ki}\, \RHO\, \MAT E_{j\ell} \Bigg) ~.
\end{equation}
For future reference, we note further that the associated Choi matrix is given by
\begin{equation} \begin{split}
\EMB{Choi}\big(\M[0.1]\EMB{\ALG S}\M[0.1]\big) ~\equiv~
& \sum_{i,j,k,\ell=0}^{N-1} \sum_{\;m,n=0}^{N-1}\, s^{ij}_{k\ell}\; \big( \MAT E_{mn} \MAT E_{\ell j} \otimes \MAT E_{ki} \MAT E_{mn} \big) \\ =~
& \sum_{i,j,k,\ell=0}^{N-1} s^{ij}_{k\ell}\; \big( \MAT E_{ij} \otimes \MAT E_{k\ell} \big) ~=\, \sum_{i,j,k,\ell=0}^{N-1} s^{ij}_{k\ell}\; (\MAT e_i \otimes \MAT e_k)\, {(\MAT e_j \otimes \MAT e_\ell)}^\dag ~.
\end{split} \end{equation}
This shows that while the representative supermatrix $\EMB{\ALG S}$ in this basis is formed by identically ordering the upper and lower index pairs of $s^{ij}_{k\ell}$ and using the resulting list as the row and column indices, the Choi matrix is obtained by ordering the right and left index pairs and using the result as the row and column indices, respectively.

\begin{LEMMA} \label{thm:trpr}
A superoperator $\ALG S$ with representative matrix $\EMB{\ALG S} = \big[ s^{ij}_{k\ell}\big]_{k,\ell;\,i,j\,=\,0}^{\,N\,-\,1}$ versus a Hilbert space basis $\{ \MAT e_i \}_{i=0}^{N-1}$ preserves the trace of its operands if and only if
\begin{equation}
\COL^\dag(\MAT I)\, \EMB{\ALG S}\, (\MAT e_i \otimes \MAT e_j) ~=~ \sum_{k=0}^{N-1}\, s^{ij}_{kk} ~=~ \delta^{ij} \quad\text{for}~0\le i,j < N ~,
\end{equation}
where $\delta^{ij}$ is a Kronecker delta.
\end{LEMMA}
\PROOF The usual trace-preservation condition can be written as
\begin{align}
\MAT I ~=~ \notag
& \sum_{i,j,k,\ell=0}^{N-1} s^{ij}_{k\ell}\; \MAT E_{j\ell}\, \MAT E_{ki} ~=~ \sum_{i,j,k=0}^{N-1} s^{ij}_{kk}\; \MAT E_{ji} \\ =~
& \sum_{i,j=0}^{N-1}\, \Bigg( \sum_{k=0}^{N-1}\, {(\MAT e_k \otimes \MAT e_k)}^\dag\, \EMB{\ALG S}\, (\MAT e_j \otimes \MAT e_i) \Bigg) \MAT E_{ji} \\ =~ \notag
& \sum_{i,j=0}^{N-1}\, \Big( \COL^\dag(\MAT I)\, \EMB{\ALG S}\, (\MAT e_i \otimes \MAT e_j) \Big) \MAT E_{ij} ~,
\end{align}
which is equivalent to the stated conditions. QED

The Lemma can be stated more succinctly by saying that $\COL(\MAT I)$ is a left-eigenvector of $\EMB{\ALG S}$ with eigenvalue $1$. We note that for another important class of superoperators, namely the identity preserving or \emph{unital} superoperators, the operator sum representations satisfy $I = \ALG S(I) = \sum_{m,n=0}^{N^2-1} \tau_{mn} T_{\X[1.35]m} T_n^\dag$, may also be characterized in terms of their supermatrix representations by $\sum_{i=0}^{N-1} s^{ii}_{k\ell\,} = (\MAT e_k \otimes \MAT e_\ell)^{\dag\,} \EMB{\ALG S}\, \COL(\MAT I) = \delta_{k\ell}$, i.e.~$\COL(\MAT I)$ is a right-eigenvector of $\EMB{\ALG S}$ with eigenvalue $1$. If $\ALG S = \ALG S(t)$ is a unital QDS, it is easily seen that the corresponding Lindblad operators must be normal (or commute with the adjoints).

Returning now to the problem of deriving a Lindblad representation for a QDS $\ALG S(t)$ from a matrix exponential representation $\EMB{\ALG S}(t) = \EMB{Exp}(-\ALG F\,t)$ thereof, the obvious way to proceed, given the results of the previous section, is to simply differentiate it:
\begin{equation} \begin{split}
& \partial_t\, \EMB{Exp}(-\EMB{\ALG F}\,t) \big|_{t=0} ~=~ -\EMB{\ALG F} ~\equiv~ -\! \sum_{i,j,k,\ell=0}^{N-1} f^{ij}_{k\ell}\; (\MAT E_{\ell j} \otimes \MAT E_{ki}) \\ =~
& \partial_t \sum_{i,j,k,\ell=0}^{N-1} s^{ij}_{k\ell}\; (\MAT E_{\ell j} \otimes \MAT E_{ki}) \Big|_{t=0} ~\equiv~ \sum_{i,j,k,\ell=0}^{N-1} \dot s^{ij}_{k\ell}\; (\MAT E_{\ell j} \otimes \MAT E_{ki})
\end{split} \end{equation}
(note that the generator is actually time-independent). Differentiation of our trace-preservation condition similarly yields $\sum_{k=0}^{N-1} \dot s^{ij}_{kk} = -\sum_{k=0}^{N-1} f^{ij}_{kk} = 0$ for all $0 \le i,j < N$, and hence
\begin{equation} \begin{split}
\label{eq:ideally}
\dot \RHO(t) ~=~
& \sum_{i,j,k,\ell=0}^{N-1} \dot s^{ij}_{k\ell}\; \MAT E_{ki}\, \RHO(t)\, \MAT E_{j\ell} ~=~ -\sum_{i,j,k,\ell=0}^{N-1} f^{ij}_{k\ell}\; \MAT E_{ki}\, \RHO(t)\, \MAT E_{j\ell} \\ =~
& \dot \RHO(t) \,-\, \HALF \sum_{i,j,k=0}^{N-1} \dot s^{ij}_{kk} \Big( \MAT E_{ji}\, \RHO \,+\, \RHO\, \MAT E_{ji} \Big) \\ =~
& \sum_{i,j,k,\ell=0}^{N-1} \dot s^{ij}_{k\ell}\; \Big(\MAT E_{ki}\, \RHO(t)\, \MAT E_{j\ell} \,-\, \HALF \big( \MAT E_{j\ell}\, \MAT E_{ki} \RHO \,+\, \RHO\, \MAT E_{j\ell}\, \MAT E_{ki} \big)\! \Big) ~.
\end{split} \end{equation}
Thus we could in principle obtain a canonical Lindblad representation for $\dot\RHO(t)$ simply by diagonalizing the (time-independent) Choi matrix of the generator $\big[\dot s^{ij}_{k\ell} \big]_{i,k;\,j,\ell\,=\,0}^{\,N\,-\,1} = \big[ \sum_{m,n=0}^{N-1} \varphi^m_n \, u^{im}_{kn} \, \bar u^{jm}_{\ell n} \big]_{i,k;\,j,\ell\,=\,0\,}^{\,N\,-\,1}$, and letting the Lindblad operators be defined by the matrices
\begin{equation}
\MAT L^m_n ~\equiv~ \sqrt{\varphi^m_n}\, \sum_{i,k=0}^{N-1}\, u^{im}_{kn}\, \MAT E_{ik}
\end{equation}
--- \emph{providing} that the eigenvalues $\varphi^m_n \ge0$ for all $0\le m,n < N$.
But then our trace-preservation condition for $\dot\RHO$ implies
\begin{equation} \begin{split}
\sum_{m,n=0}^{N-1}\, \big( \MAT L^m_n \big)^{\!\dag\,} \MAT L^m_n ~=~
& \sum_{m,n=0}^{N-1}\, \varphi^m_n\; \sum_{i,j,k,\ell=0}^{N-1} u^{im}_{kn}\, \bar u^{jm}_{\ell n} \,\MAT E_{j\ell}\, \MAT E_{ki} \\ =~
& \sum_{i,j,k=0}^{N-1} \sum_{m,n=0}^{N-1} \Bigg( \varphi^m_n\; u^{im}_{kn} \bar u^{jm}_{kn} \Bigg) \MAT E_{ji} ~=~ \sum_{i,j,k=0}^{N-1} \dot s^{ij}_{kk} \MAT E_{ji} ~=~ \MAT 0 ~,
\end{split} \end{equation}
which contradicts the fact that a nontrivial sum of positive semidefinite matrices cannot vanish. This result is easily shown to be independent of the choice of matrix basis.

It follows that there must be some redundance in our choice of coefficients in any nondiagonal Lindblad-type equation of the form given in Eq.~(\ref{eq:ideally}). Moreover, such an equation, by its very form, is assured of preserving the trace ($\partial_t\, \TR(\RHO(t)) = \TR(\dot\RHO(t)) = 0$), so that the trace-preservation condition satisfied by the derivatives of the coefficients in an operator sum representation is not needed. Our problem is to find a way to modify the matrix of coefficients $\big[\dot s^{ij}_{k\ell} \big]_{m,n=0}^{N-1}$, while preserving the underlying mapping $\RHO \mapsto \dot\RHO$,  such that the result is positive semidefinite and so can be diagonalized to obtain a canonical Lindbladian. Because any Lindblad operator of the form $\MAT L = \alpha\,\MAT I$ with $\alpha \in \FLD C$ adds nothing to $\dot\RHO$, we shall seek to eliminate the corresponding degree of freedom from the coefficients.

\begin{LEMMA}
In any quantum dynamical semigroup with exponential representation $\ALG S(t) = Exp(-\ALG F\,t)$, the generator's matrix $\EMB{\ALG F}$ versus a Hilbert space basis satisfies
\begin{equation}
\label{eq:isneg}
\COL^\dag(\MAT I)\, \EMB{Choi}(-\EMB{\ALG F})\, \COL(\MAT I) ~<~ 0 ~.
\end{equation}
If the generator is of the form $\ALG F = \ALG G + \imath\ALG H$ where $\ALG H$ is a commutation superoperator and $\big\langle \ALG G, \ALG C \big\rangle \equiv \TR( \ALG G^\dag \ALG C ) = 0$ for any commutation superoperator $\ALG C$, then the corresponding matrix projection satisfies
\begin{equation}
\label{eq:nocom}
\EMB{\ALG P}^{\MAT I}\, \EMB{Choi}(\EMB{\ALG G})\,  \EMB{\ALG P}^{\MAT I}\, ~=~ \EMB{\ALG P}^{\MAT I}\, \EMB{Choi}(\EMB{\ALG F})\, \EMB{\ALG P}^{\MAT I} \qquad\big( \EMB{\ALG P}^{\MAT I} ~\equiv~ \MAT I \otimes \MAT I \,-\, \COL(\MAT I)\, \COL^\dag(\MAT I) / N \big) ~.
\end{equation}
\end{LEMMA}
\PROOF To prove Eq.~(\ref{eq:isneg}), we first observe that
\begin{equation} \begin{split}
& \COL^\dag(\MAT I)\, \EMB{Choi}(-\EMB{\ALG F})\, \COL(\MAT I) ~=~ \!-\! \sum_{m,n=0}^{N-1} {(\MAT e_m \otimes \MAT e_m)}^\dag \Bigg( \sum_{i,j,k,\ell=0}^{N-1} f^{ij}_{k\ell}\, (\MAT E_{ij} \otimes \MAT E_{k\ell})\! \Bigg) (\MAT e_n \otimes \MAT e_n) \\ \M[-0.75] =~
& -\!\sum_{m,n=0}^{N-1} \Bigg( \sum_{i,j,k,\ell=0}^{N-1} \Big( f^{ij}_{k\ell}\, \big( (\MAT e_m^\dag \MAT E_{ij\,} \MAT e_n) \otimes (\MAT e_m^\dag \MAT E_{k\ell\,} \MAT e_n) \big)\! \Big)\! \Bigg) ~=~ -\!\sum_{m,n=0}^{N-1}\, f^{mn}_{mn} ~=~ -\TR( \EMB{\ALG F} ) ~.
\end{split} \end{equation}
Since $\EMB{Choi}(-\EMB{\ALG F})$ is Hermitian, this quantity is real, and since $\ALG S(t)$ is bounded, the eigenvalues of $-\EMB{\ALG F}$ must all have negative real parts, so that $-\TR(\EMB{\ALG F}) < 0$.

To prove Eq.~(\ref{eq:nocom}), we first note that it is sufficient to prove this for the commutation superoperator of an arbitrary elementary matrix $\MAT E_{ij}$, and transform its generating supermatrix to the corresponding Choi matrix:
\begin{equation} \begin{split}
\label{eq:compro}
\EMB{Choi}\big( \MAT E_{ij} \otimes \MAT I \,-\,\MAT I \otimes \MAT E_{ji}\big) ~=~
& \sum_{k,\ell=0}^{N-1}\, \big( (\MAT E_{k\ell}\, \MAT E_{ij}) \otimes \MAT E_{k\ell} \,-\, \MAT E_{k\ell} \otimes (\MAT E_{ji} \MAT E_{k\ell}) \big) \\ =~
& \sum_{k=0}^{N-1}\, \big( \MAT E_{kj} \otimes \MAT E_{ki} \,-\, \MAT E_{ik} \otimes \MAT E_{jk} \big) ~.
\end{split} \end{equation}
Plugging the second term into the projection now yields
\begin{equation} \begin{split}
\EMB{\ALG P}^{\MAT I} \Bigg( \sum_{k=0}^{N-1}\, \MAT E_{ik} \otimes \MAT E_{jk} \Bigg) \EMB{\ALG P}^{\MAT I} ~=~
& \sum_{k=0}^{N-1}\, \MAT E_{ik} \otimes \MAT E_{jk} \,-\, \frac{\delta_{ij}}N\, \COL(\MAT I)\, \sum_{k=0}^{N-1}\, {(\MAT e_k \otimes \MAT e_k)}^\dag \\
& \quad -\, (\MAT e_i \otimes \MAT e_j) \, \COL^\dag(\MAT I) \,+\, \frac{\delta_{ij}}N\, \COL(\MAT I)\, \COL^\dag(\MAT I) ~.
\end{split} \end{equation}
Since the first and third terms as well as the second and fourth terms on the right-hand side differ only in sign, this projection vanishes identically. A similar calculation shows that the projection of the first term on the right-hand side of Eq.~(\ref{eq:compro}) likewise vanishes, establishing the Lemma. QED

Henceforth, we take $\ALG G = \ALG F - \imath\ALG H$ where $\ALG H$ is the commutator part of $\ALG F$, and let $g^{ij}_{k\ell}$ be the corresponding array of coefficients. A final technical Lemma will be needed to prove the first real result in this section.
\begin{LEMMA}
\label{thm:commcondcoef}
If $\big\langle \ALG G, \ALG C \big\rangle \equiv \TR( \ALG G^\dag \ALG C ) = 0$ for every commutation superoperator $\ALG C$ as above, then the coefficients $g^{ij}_{k\ell}$ of any supermatrix representation $\EMB{\ALG G}$ satisfy
\begin{equation}
\sum_{k=0}^{N-1}\, g^{kn}_{km} ~=~ \sum_{\ell=0}^{N-1}\, g^{m\ell}_{n\ell}
\end{equation}
for all $0 \le m, n < N$.
\end{LEMMA}
\PROOF The proof is by direct computation:
\begin{alignat}{2}
& 0 \M & =\M &
\TR\Bigg( \big( \MAT E_{nm} \otimes \MAT I \,-\, \MAT I \otimes \MAT E_{mn} \big) \sum_{i,j,k,\ell=0}^{N-1} g_{k\ell}^{\,ij}\, \big( \MAT E_{\ell j} \otimes \MAT E_{ki} \big) \Bigg)
\notag \\ && =\M &
\TR\Bigg( \sum_{i,j,k,\ell=0}^{N-1} g_{k\ell}^{\,ij}\, \big( (\MAT E_{nm}\, \MAT E_{\ell j}) \otimes \MAT E_{ki} \,-\, \MAT E_{\ell j} \otimes (\MAT E_{mn}\, \MAT E_{ki}) \big) \Bigg)
\notag \\[-\baselineskip] \\ \notag \implies\qquad &&&
\TR\Bigg( \sum_{i,j,k=0}^{N-1} g_{km}^{\,ij}\, \big( \MAT E_{nj} \otimes \MAT E_{ki} \big) \Bigg) ~=\M[0.66] \TR\Bigg( \sum_{i,j,\ell=0}^{N-1} g_{n\ell}^{\,ij}\, \big( \MAT E_{\ell j} \otimes \MAT E_{mi} \big) \Bigg)
\\ \notag \implies\qquad &&& \M[0.5]
\sum_{i,j,k=0}^{N-1} g_{km}^{\;ij}\; \TR\big( \MAT E_{nj} \big)\, \TR\big( \MAT E_{ki} \big) ~=~ \sum_{i,j,\ell=0}^{N-1} g_{n\ell}^{ij}\; \TR\big( \MAT E_{\ell j} \big)\, \TR\big( \MAT E_{mi} \big)
\\ \notag \implies\qquad &&& \M[4.5]
\sum_{i,j,k=0}^{N-1} g_{km}^{\;ij}\; \delta_{nj}\, \delta_{ki} ~=~ \sum_{i,j,\ell=0}^{N-1} g_{n\ell}^{ij}\; \delta_{\ell j}\, \delta_{mi}
\end{alignat}
QED\bigskip\break
This Lemma may be paraphrased by saying that the ``partial trace'' (or contraction) of $\EMB{\ALG G}$ with respect to either its left or right Kronecker factors are the transposes of one another.

\begin{PROPOSITION}
\label{thm:lindblad}
Let $\ALG S(t) = Exp(-\ALG F\,t)$ be a quantum dynamical semigroup with $\ALG F = \ALG G \M[0.05]+\M[0.05] \imath\ALG H$ as above. Then if their supermatrices versus a Hilbert space basis are $\EMB{\ALG F} = \big[ f^{\,ij}_{k\ell} \big]_{k,\ell;\,i,j\,=\,0}^{\,N\,-\,1}\,$, $\EMB{\ALG G} = \big[ g^{\,ij}_{k\ell} \big]_{k,\ell;\,i,j\,=\,0}^{\,N\,-\,1}$ and $\EMB{\ALG H} = \MAT I \otimes \MAT H - \OL{\MAT H} \otimes \MAT I$, we have
\begin{align}
\label{eq:lindblad}
\M[-2] \dot\RHO(t) ~\equiv~ \notag
& -\! \sum_{i,j,k,\ell=0}^{N-1} f^{\,ij}_{k\ell}\; \MAT E_{ki}\, \RHO(t)\, \MAT E_{j\ell} ~\equiv\, \imath\big[ \RHO(t),\, \MAT H \big]\, - \sum_{i,j,k,\ell=0}^{N-1} g^{\,ij}_{k\ell}\; \MAT E_{ki}\, \RHO(t)\, \MAT E_{j\ell} \\ =~
& \imath\big[ \RHO(t),\, \MAT H \big] \,-\, \frac12 \sum_{i,j,k,\ell=0}^{N-1} \check g^{\,ij}_{k\ell}\, \Big( 2\, \MAT E_{ki}\, \RHO(t)\, \MAT E_{j\ell} \,-\, \MAT E_{j\ell}\, \MAT E_{ki}\, \RHO(t) \,-\, \RHO(t)\, \MAT E_{j\ell}\, \MAT E_{ki} \Big) \\ =~ \notag
& \imath\big[ \RHO(t),\, \MAT H \big] \,-\, \sum_{i,j=0}^{N-1} \Bigg( \sum_{k,\ell=0}^{N-1} \, \check g^{\,ij}_{k\ell}\, \MAT E_{ki}\, \RHO(t)\, \MAT E_{j\ell} \,-\, \frac12 \sum_{k=0}^{N-1}\, \check g^{\,ij}_{kk}\, \Big( \MAT E_{ji}\, \RHO(t) \,+\, \RHO(t)\, \MAT E_{ji} \Big) \! \Bigg)
\end{align}
where $\big[ \check g^{\,ij}_{k\ell} \big]_{k,\ell;\,i,j\,=\,0}^{\,N\,-\,1}$ are the coefficients of the supermatrix
\begin{equation}
\check{\EMB{\ALG G}} ~\equiv~ \EMB{Choi}( \EMB{\ALG P}^{\MAT I}\, \EMB{Choi}( \EMB{\ALG G} )\, \EMB{\ALG P}^{\MAT I} ) ~=~ \EMB{Choi}( \EMB{\ALG P}^{\MAT I}\, \EMB{Choi}( \EMB{\ALG F} )\, \EMB{\ALG P}^{\MAT I} ) ~.
\end{equation}
\end{PROPOSITION}
\PROOF Note that $\ALG H$ occurs on both sides of Eq.~(\ref{eq:lindblad}), so we can just ignore it (i.e.~set $\MAT H = \MAT 0$) in the proof. Since $\COL(\MAT I)\COL^\dag(\MAT I) = \sum_{m,n=0}^{N-1} \MAT E_{mn} \otimes \MAT E_{mn}$, we find that: $\EMB{\ALG P}^{\MAT I}\,\EMB{Choi}(\EMB{\ALG G})\,\EMB{\ALG P}^{\MAT I}$
\begin{align}
\label{eq:bigmess}
=~ \notag
& \begin{aligned}[t]
\sum_{i,j,k,\ell=0}^{N-1} g_{k\ell}^{ij}\, \Bigg( \MAT E_{ij} \otimes \MAT E_{k\ell} ~-~
& \frac1N\, \sum_{m,n=0}^{N-1}\, \big( \MAT E_{mn} \MAT E_{ij} \otimes \MAT E_{mn} \MAT E_{k\ell} + \MAT E_{ij} \MAT E_{mn} \otimes \MAT E_{k\ell} \MAT E_{mn} \big) \\ +~
& \frac1{N^2}\, \sum_{m,n,p,q=0}^{N-1} \MAT E_{mn} \MAT E_{ij} \MAT E_{pq} \otimes \MAT E_{mn} \MAT E_{k\ell} \MAT E_{pq} \Bigg)
\end{aligned}
\\ =~ & \begin{aligned}[t]
\sum_{i,j,k,\ell=0}^{N-1} g_{k\ell}^{ij}\, \Bigg( \MAT E_{ij} \otimes \MAT E_{k\ell}
~ &-~\frac1N\, \bigg( \delta^i_k\, \sum_{m=0}^{N-1}\, \MAT E_{mj} \otimes \MAT E_{m\ell}
\,+\, \delta^j_\ell\, \sum_{n=0}^{N-1}\, \MAT E_{in} \otimes \MAT E_{kn} \bigg)
\\&+~ \frac1{N^2}\; \delta_k^i\, \delta_\ell^j \sum_{m,n=0}^{N-1} \MAT E_{mn}
\otimes \MAT E_{mn} \Bigg)
\end{aligned}
\\ =~ \notag
& \sum_{i,j,k,\ell=0}^{N-1} \bigg( g_{k\ell}^{ij} ~-~ \frac{\delta_k^i}N\,
\sum_{m=0}^{N-1}\, g_{m\ell}^{mj} ~-~ \frac{\delta_\ell^j}N\,\sum_{n=0}^{N-1}\,
g_{kn}^{in} ~+~ \frac{\delta_k^i\delta_\ell^j}{N^2} \sum_{m,n=0}^{N-1}\,
g_{mn}^{mn} \bigg)\, \MAT E_{ij} \otimes \MAT E_{k\ell}
\\ \equiv~ \notag
& \sum_{i,j,k,\ell=0}^{N-1} \check g_{k\ell}^{ij}\; \MAT E_{ij} \otimes \MAT E_{k\ell} ~.
\end{align}
Equation (\ref{eq:bigmess}) thus determines the projected coefficients $\check g^{\,ij}_{k\ell}$ in terms of the original coefficients, and if we compute the Lindbladian versus a Hilbert space basis using the projected coefficients as in the last line of Eq.~(\ref{eq:lindblad}), we get
\begin{equation}
\label{eq:whatuwant}
\ALG L(\RHO) ~\equiv~ \sum_{i,j=0}^{N-1} \Bigg( \sum_{k,\ell=0}^{N-1} \, \check g^{\,ij}_{k\ell}\, \MAT E_{ki}\, \RHO(t)\, \MAT E_{j\ell} \,-\, \frac12 \sum_{k=0}^{N-1}\, \check g^{\,ij}_{kk}\, \Big( \MAT E_{ji}\, \RHO(t) \,+\, \RHO(t)\, \MAT E_{ji} \Big) \! \Bigg) ~.
\end{equation}
The supermatrix representation of the first operator sum in this equation can be further simplified as follows:
\begin{equation} \begin{split}
\label{eq:rhsoflin}
\sum_{i,j,k,\ell=0}^{N-1} \check g_{k\ell}^{ij}\; \MAT E_{\ell j} \otimes \MAT E_{ki} ~=~
& \begin{aligned}[t]
& \sum_{i,j,k,\ell=0}^{N-1} g_{k\ell}^{\,ij}\, \MAT E_{\ell j} \otimes \MAT E_{ki} \,-\, \sum_{j,\ell=0}^{N-1}\, \bigg( \frac1N \sum_{m=0}^{N-1}\, g^{mj}_{m\ell} \bigg) \MAT E_{\ell j} \otimes \MAT I
\\ &
\,-\, \sum_{i,k=0}^{N-1}\, \bigg( \frac1N \sum_{n=0}^{N-1}\, g^{\,in}_{kn} \bigg) \MAT I \otimes \MAT E_{ki} \,+\,\bigg( \frac1{N^2}\! \sum_{m,n=0}^{N-1}\, g_{mn}^{mn} \bigg) \MAT I \otimes \MAT I ~.
\end{aligned}
\end{split} \end{equation}
Similarly, by Eq.~(\ref{eq:bigmess}) the supermatrix representation of the second operator sum in Eq.~(\ref{eq:whatuwant}) simplifies to
\pagebreak[2]
\begin{equation} \begin{split}
\label{eq:lhsoflin}
& \frac12 \sum_{i,j,k,\ell=0}^{N-1} \delta_{k\ell}\; \check g^{\,ij}_{k\ell}\, \big( \MAT E_{ji} \otimes \MAT I \,+\, \MAT I \otimes \MAT E_{ij} \big) \\ =~
& \frac12\, \sum_{i,j=0}^{N-1} \Bigg( \sum_{k=0}^{N-1}\, g_{kk}^{ij} \,-\, \frac1N\, \sum_{m=0}^{N-1}\, g_{mi}^{mj} \,-\, \frac1N\, \sum_{n=0}^{N-1}\, g_{jn}^{\,in} \Bigg) \big( \MAT E_{ij} \otimes \MAT I + \MAT I \otimes \MAT E_{ji} \big) \\
& \M[10] +~\bigg( \frac1{N^2}\, \sum_{m,n=0}^{N-1}\, g_{mn}^{mn} \bigg) \MAT I \otimes \MAT I ~.
\end{split} \end{equation}
Taking into account the difference in the signs of the operator sums in Eq.~(\ref{eq:whatuwant}), the last terms on the right-hand sides of Eqs.~(\ref{eq:rhsoflin}) and (\ref{eq:lhsoflin}) clearly cancel, while the first summation on the right-hand Eq.~(\ref{eq:lhsoflin}) vanishes by our trace preservation condition (cf.~Lemma \ref{thm:trpr}). The remaining terms on the right-hand side of Eq.~(\ref{eq:lhsoflin}) can be rearranged using the symmetries of the summations proved in Lemma \ref{thm:commcondcoef}, as follows:
\begin{align}
& -\frac1{2N}\, \sum_{i,j=0}^{N-1} \Bigg( \sum_{m=0}^{N-1}\, g_{mi}^{mj} \,+\, \sum_{n=0}^{N-1}\, g_{jn}^{\,in} \Bigg) \big( \MAT E_{ij} \otimes \MAT I + \MAT I \otimes \MAT E_{ji} \big) \notag \\ =~
& \begin{aligned}[t]
-\frac1{2N}\, \sum_{i,j=0}^{N-1}\, \Bigg(
& \bigg( \sum_{m=0}^{N-1}\, g_{mi}^{mj} \bigg)\, \MAT E_{ij} \otimes \MAT I \,+\, \bigg( \sum_{m=0}^{N-1}\,  g_{jm}^{\,im} \bigg)\, \MAT I \otimes \MAT E_{ji} \\
& +\, \bigg( \sum_{n=0}^{N-1}\, g_{ni}^{\,nj} \bigg)\, \MAT E_{ij} \otimes \MAT I \,+\, \bigg( \sum_{n=0}^{N-1}\, g_{jn}^{\,in} \bigg)\, \MAT I \otimes \MAT E_{ji} \Bigg)
\end{aligned}
\\ \notag =~
& -\frac1N\, \sum_{i,j=0}^{N-1}\, \Bigg( \bigg( \sum_{m=0}^{N-1}\, g_{mi}^{mj} \bigg)\, \MAT E_{ij} \otimes \MAT I \,+\, \bigg( \sum_{n=0}^{N-1}\, g_{jn}^{\,in} \bigg)\, \MAT I \otimes \MAT E_{ji} \Bigg) ~.
\end{align}
It is now apparent that these terms cancel with the second and third terms in Eq.~(\ref{eq:rhsoflin}) after a change of dummy indices, leaving only its first term behind. QED

Thus, roughly speaking, the subtraction of the $(L^\dag L \rho + \rho L^\dag L)/2$ terms from the $L \rho L^\dag$ terms of the Lindbladian simply ensures the trace of $\dot\rho$ still vanishes after projecting out the commutator and identity superoperator parts of the derivative of the corresponding Kraus operator sum. It remains to be shown that the Choi matrix of the operator sum is positive semidefinite if and only if the Choi matrix of the projection of its derivative is positive semidefinite. For the sake of completeness, we first prove the following (well-known) result, using only the techniques developed above.
\begin{LEMMA}
\label{thm:cpospro}
The composition of two completely positive superoperators $\ALG A \circ \ALG B$ is again completely positive.
\end{LEMMA}
\PROOF Let $\EMB{\ALG U}\,\DMAT(\EMB\alpha)\,\EMB{\ALG U}^\dag$ and $\EMB{\ALG V}\,\DMAT(\EMB\beta)\,\EMB{\ALG V}^\dag$ be the eigenvalue decompositions of the supermatrices $\EMB{Choi}(\EMB{\ALG A})$ and $\EMB{Choi}(\EMB{\ALG B})$ respectively, and consider the Choi matrix of their product, namely
\begin{align*}
\refstepcounter{equation}
\EMB{Choi}\big( \EMB{\ALG A}\, \EMB{\ALG B} \big) ~
=~ &
\sum_{i,j,k,\ell=0}^{N-1} \Bigg(
\sum_{m,n=0}^{N-1}\, a^{mn}_{k\ell} b^{ij}_{mn}
\Bigg)\, \MAT E_{ij} \otimes \MAT E_{k\ell}
\\ =~ &
\sum_{i,j,k,\ell=0}^{N-1} \Bigg( \sum_{m,n=0}^{N-1} \bigg(
\sum_{p,q=0}^{N-1}\, u^{mp}_{kq} \bar u^{np}_{\ell q} \alpha^p_q
\bigg) \bigg(
\sum_{r,s=0}^{N-1}\, v^{ir}_{ms} \bar v^{jr}_{ns} \beta^{\,r}_s
\bigg)\, \MAT E_{ij} \otimes \MAT E_{k\ell} \Bigg)
\\ =~ &
\sum_{i,j,k,\ell=0}^{N-1} \Bigg(
\sum_{p,q,r,s=0}^{N-1} \alpha^p_q\, \beta^{\,r}_s
\bigg( \sum_{m=0}^{N-1}\, u^{mp}_{kq} v^{ir}_{ms} \bigg)
\bigg( \sum_{n=0}^{N-1}\, \bar u^{np}_{\ell q} \bar v^{jr}_{ns} \bigg)
\, \MAT E_{ij} \otimes \MAT E_{k\ell} \Bigg) \qquad\qquad(\theequation)
\\ =~ &
\sum_{p,q,r,s=0}^{N-1} \alpha^p_q\, \beta^{\,r}_s
\Bigg( \sum_{i,k=0}^{N-1}\, (\MAT e_i \otimes \MAT e_k)
\bigg( \sum_{m=0}^{N-1}\, u^{mp}_{kq} v^{ir}_{ms} \bigg) \Bigg)
\Bigg( \sum_{j,\ell=0}^{N-1}\, (\MAT e_j \otimes \MAT e_\ell)^\dag
\bigg( \sum_{n=0}^{N-1}\, \bar u^{np}_{\ell q} \bar v^{jr}_{ns} \bigg)
\Bigg)
\\ \equiv~ & 
\sum_{p,q,r,s=0}^{N-1} \alpha^p_q\, \beta^{\,r}_s\,
\EMB x^{pr}_{qs}\, \big( \EMB x^{pr}_{qs} \big)^\dag ~.
\end{align*}
Such a sum of positive semidefinite matrices
(Hermitian dyads, in this case) is always again
positive semidefinite, proving the Lemma. QED

\begin{THEOREM}
The integral of a Lindbladian yields a quantum dynamical semigroup, and conversely, and the derivative of any quantum dynamical semigroup can be placed in canonical Lindblad form.
\end{THEOREM}
\PROOF Given any Kraus operator sum for a quantum dynamical semigroup $\ALG S(t)$, we know that its time-derivative will be equal to the result of applying a fixed generator $-\ALG F$ to the density operator $\rho(t)$ at any given $t\ge0$. Integration of a matrix representation thus yields $\EMB{\ALG S}(t) = \EMB{Exp}(-\EMB{\ALG F}\,t)$, and for a sufficiently small $\delta t > 0$ this exponential may be approximated arbitrarily closely by
\begin{equation}
\EMB{Exp}\big(\!-\EMB{\ALG F}t\big) ~\approx~
\EMB{\ALG I} \,-\, \EMB{\ALG F}\, \delta t  ~+~ O\big( (\delta t)^2 \big)
~=~ \EMB{\ALG I} \,-\, \EMB{\ALG G}\, \delta t \,-\,
\imath\EMB{\ALG H}\, \delta t ~+~ O\big( (\delta t)^2 \big) ~,
\end{equation}
where $\EMB{\ALG I} \equiv \MAT I\otimes\MAT I$ and $\imath\EMB{\ALG H}$ denotes the commutator part of $\EMB{\ALG F}$. Since $\ALG S(t)$ is completely positive, any Choi matrix for it must be positive semidefinite, and so must any projection thereof, in particular,
\begin{equation}
\EMB{\ALG P}^{\MAT I}\,\EMB{Choi}\big( \EMB{\ALG I} \,-\, \EMB{\ALG F}\, \delta t \big)\,\EMB{\ALG P}^{\MAT I} ~=~
-\EMB{\ALG P}^{\MAT I}\, \EMB{Choi}\big(\EMB{\ALG G} \big)\, \EMB{\ALG P}^{\MAT I}\, \delta t ~\equiv~
- \EMB{Choi}\big( \check{\EMB{\ALG G}} \big)\, \delta t ~.
\end{equation}
This allows $-\check{\ALG G}$ and hence also its sum with $-\imath\ALG H$ to be placed in canonical Lindblad form, which by Proposition \ref{thm:lindblad} must have the same action on any $\rho$ as the differential superoperator $-\ALG F$.

Conversely, suppose that a given superoperator $\ALG F = \ALG G + \imath\ALG H$ can be placed in canonical Lindblad form,
\begin{equation}
-\ALG F(\rho) ~=~\ALG L(\rho) ~\equiv\; -\imath\ALG H(\rho) \,+\, \sum_{m=1}^M\, \Big( L_m\, \rho\, L_m^\dag - \HALF L_m^\dag L_m\, \rho - \HALF\M[0.05] \rho\, L_m^\dag L_m \Big)
\end{equation}
where $\ALG H(\rho) \equiv [H, \rho\M[0.05]]$ for the commutator part of $\ALG F$. In terms of a matrix representation $\MAT H$, $\MAT L_m$ of these operators, this is equivalent to
\begin{equation}
\label{eq:recon}
-\EMB{\ALG F} ~=\; - \EMB{\ALG G} - \imath\EMB{\ALG H} ~\equiv~ \begin{aligned}[t]
\sum_{m=1}^M \Big( \OL{\MAT L}_m \otimes \MAT L_m - \HALF\, \MAT I \otimes \MAT L_m^\dag \MAT L_{\X[1.35]m} - \HALF\, \OL{\MAT L}_m^\dag \OL{\MAT L}_{\X[1.35]m} \otimes \MAT I \Big) & \\
 +\; \imath \big(\, \OL{\!\MAT H\!} \otimes \MAT I - \MAT I \otimes \MAT H \big) & ~. \end{aligned}
\end{equation}
Then over a sufficiently small time interval $\delta t$, the exponential (integral) can be approximated arbitrarily closely by the product of the exponentials
\begin{equation} \begin{split}
\EMB{Exp}\big(\! -\! \delta t\, \EMB{\ALG F} \big) ~\approx~ &
\EMB{Exp}\bigg(\!\! -\! \HALF\delta t \sum_{m=1}^M \Big( \MAT I \otimes \MAT L_m^\dag \MAT L_{\X[1.35]m} + \OL{\MAT L}_m^\dag \OL{\MAT L}_{\X[1.35]m} \otimes \MAT I \Big)\! \bigg)\, \cdots \\ &
\cdots\, \EMB{Exp}\bigg(\! \delta t \sum_{m=1}^M\, \OL{\MAT L}_m \otimes \MAT L \bigg)\, \EMB{Exp}\big(\! -\! \delta t\, \imath\EMB{\ALG H}\, \big) ~+~ O\big((\delta t)^2\big) \\
~\equiv~ & \EMB{\ALG A}(\delta t)\, \EMB{\ALG B}(\delta t)\, \EMB{\ALG C}(\delta t)
~+~ O\big((\delta t)^2\big) ~.
\label{eq:approx}
\end{split} \end{equation}

Since the two types of terms in the argument to the first exponential commute, it evaluates to a Kronecker product, namely
\begin{equation} \begin{split}
\EMB{\ALG A}(\delta t) ~\equiv~ & \EMB{Exp}\bigg(\!\! - \HALF\delta t \sum_{m=1}^M \Big( \MAT I \otimes \MAT L_m^\dag \MAT L_{\X[1.35]m} + \OL{\MAT L}_m^\dag \OL{\MAT L}_{\X[1.35]m} \otimes \MAT I \Big)\! \bigg) \\ =~ &
\EMB{Exp}\bigg(\!\! - \HALF\delta t \sum_{m=1}^M\, \MAT L_m^\dag \MAT L_{\X[1.35]m} \!\bigg) \otimes\, \EMB{Exp}\bigg(\!\! - \HALF\delta t \sum_{m=1}^M\, \OL{\MAT L}_m^\dag \OL{\MAT L}_{\X[1.35]m} \!\bigg)
\end{split} \end{equation}
Thus by Lemma \ref{thm:keycor}, the corresponding Choi matrix is the dyad
\begin{equation}
\EMB{Choi}\big( \EMB{\ALG A}(\delta t) \big) ~=~ \COL\bigg(\! \EMB{Exp}\bigg(\!\! - \HALF\delta t \sum_{m=1}^M\, \MAT L_m^\dag \MAT L_{\X[1.35]m} \!\bigg) \!\bigg)\; \COL^\dag\bigg(\! \EMB{Exp}\bigg(\!\! - \HALF\delta t \sum_{m=1}^M\, \MAT L_m^\dag \MAT L_{\X[1.35]m} \!\bigg) \!\bigg) ~,
\end{equation}
which is necessarily positive semidefinite, proving that $\ALG A(t)$ is a QSD all by itself. As for the second factor in Eq.~\ref{eq:approx}, we may expand it as
\begin{equation}
\EMB{\ALG B}(\delta t) ~\equiv~ \EMB{Exp}\bigg(\! \delta t \sum_{m=1}^M\, \OL{\MAT L}_m \otimes \MAT L_m \!\bigg) ~\approx~ \EMB{\ALG I} ~+~ \delta t \sum_{m=1}^M\, \OL{\MAT L}_m \otimes \MAT L_m ~+~ O\big((\delta t)^2\big) ~.
\end{equation}
Because $\EMB{\ALG P}^\MAT I\, \EMB{Choi}(\MAT X \otimes \MAT I)\, \EMB{\ALG P}^\MAT I = \EMB{\ALG P}^\MAT I\, \EMB{Choi}(\MAT I \otimes \MAT X)\, \EMB{\ALG P}^\MAT I = \MAT 0$ for all $\MAT X \in \FLD C^{N\times N}$, the Choi matrix of the summation on the right-hand side is easily seen to be $\EMB{Choi}(\check{\EMB{\ALG G}}) \equiv \EMB{\ALG P}^\MAT I\, \EMB{Choi}(\EMB{\ALG G})\, \EMB{\ALG P}^\MAT I$, so that
\begin{equation}
\EMB{Choi}(\EMB{\ALG B}(\delta t)) ~\approx~ \COL(\MAT I)\, \COL^\dag(\MAT I) ~+~ \delta t\, \EMB{Choi}(\check{\EMB{\ALG G}}) ~+~ O\big((\delta t)^2\big) ~.
\end{equation}
The Choi matrix $\EMB{Choi}(\check{\EMB{\ALG G}}) = \sum_{m=1}^M \COL(\MAT L_m)\, \COL^\dag(\MAT L_m)$ is of course positive semidefinite, and (since $\check{\EMB{\ALG G}} = \sum_{n=1}^N \check\gamma_n (\OL{\MAT U}_n \otimes \MAT U_n)$ where $\check\gamma_n$, $\COL(\MAT U_n)$ are the eigenvalues and eigenvectors of $\EMB{Choi}(\check{\EMB{\ALG G}})$) so are the Choi matrices of all higher terms in the Taylor expansion of $\EMB{Choi}(\EMB{\ALG B}(\delta t))$, thus showing that $\ALG B(t)$ is also completely positive for all $t \ge0$. Finally, the last factor of Eq.~(\ref{eq:approx}),
\begin{equation}
\EMB{\ALG C}(\delta t) ~\equiv~ \EMB{Exp}\big(\! -\!\delta t\, \imath\EMB{\ALG H}\, \big) ~,
\end{equation}
is unitary and hence likewise corresponds to a completely positive superoperator for all time.

It now follows from Lemma \ref{thm:cpospro} that for $\delta t \ll \| \ALG G \|^{-1}$, the product of all three factors $\EMB{\ALG A}(\delta t)$, $\EMB{\ALG B}(\delta t)$, $\EMB{\ALG C}(\delta t)$ in Eq.~(\ref{eq:approx}) will be completely positive, and hence for any given $t \ge 0$ the telescoping product
\begin{equation}
\EMB{Exp}\big(\! -t\, \EMB{\ALG F}\, \big) ~\approx~
\Big(\, \underset{n~\text{times}}{\underbrace{
\EMB{\ALG A}(t/n)\, \EMB{\ALG B}(t/n)\, \EMB{\ALG C}(t/n) \cdots
\EMB{\ALG A}(t/n)\, \EMB{\ALG B}(t/n)\, \EMB{\ALG C}(t/n)
}}\,\Big)^{1/n} +~ O\big((t/n)^2\big)
\end{equation}
will also be completely positive for all $n > t / \delta t$. The Theorem now follows by noting that the set of completely positive superoperators is closed, and taking the limit as $n \rightarrow \infty$. QED

\bigskip
\section{Application to Quantum Process Tomography}
The applicability of the foregoing results to QPT derives from the following theorem, whose origins can be traced back to work by Eckart, Young and Householder \cite{EckarYoung:36a, YoungHouse:38}, and has since given rise to a field of statistical data analysis widely known as ``principal component analysis'' \cite{Joliffe:86}. The present author has proven it several times in the course of his career \cite{HaKuCr$BMB:83, GreenDeath:88, Havel:98}, and regards the following proof as the simplest.
\begin{THEOREM}
\label{thm:eyh}
Let $\MAT M \in \FLD C^{N\times N}$ be a Hermitian matrix with eigenvalue decomposition
\begin{equation}
\MAT M ~=~ \MAT U^\dag\,\EMB{\Lambda}\, \MAT U ~=~ \sum_{\ell\,=\,0}^{N-1} \lambda_{\X[1.4]\ell}\, \MAT u_{\X[1.4]\ell} \, \MAT u_\ell^\dag ~,
\end{equation}
where the eigenvalues have been sorted in nonincreasing order $\lambda_\ell \ge \lambda_{\ell+1}$ for $0 \le \ell \le N-2$. Also let $\FLD P$ denote the convex cone of positive semidefinite matrices in $\FLD C^{N\times N}$ and $\EMB{\ALG P}_{\X[1.4]\!\FLD P} (\MAT M)$ be the orthogonal projection of $\MAT M$ onto $\FLD P$ with respect to the Hilbert-Schmidt (or Frobenius) matrix norm $\|\cdot\|$, which satisfies the ``least-squares'' criterion \cite{LawsonHanson:95}
\begin{equation}
\label{eq:lsq}
\big\|\, \MAT M \,-\, \EMB{\ALG P}_{\X[1.4]\!\FLD P}(\MAT M)\, \big\|^2 ~=~ \min_{\MAT M' \in \FLD P}\, \big\|\, \MAT M \,-\, \MAT M'\, \|^2 ~.
\end{equation}
Then we have
\begin{equation}
\EMB{\ALG P}_{\X[1.4]\!\FLD P}(\MAT M) ~=~ \MAT M^\star ~\equiv~ \MAT U^\dag\,\EMB{\Lambda}^\star\, \MAT U ~=~ \sum_{\ell\,=\,0}^{N^\star-1} \lambda_{\X[1.2]\ell}^\star\, \MAT u_{\X[1.4]\ell}\, \MAT u_\ell^\dag ~,
\end{equation}
where $\EMB{\Lambda}^\star$ is the diagonal matrix of eigenvalues $\EMB{\Lambda}$ with all of its $N - N^\star$ negative eigenvalues set to zero.
\end{THEOREM}
\PROOF Any positive semidefinite $N\times N$ matrix can be written as $\MAT X\MAT X^\dag$, where $\MAT X \in \FLD C^{N\times N'}$ and $N'$ is its rank. It follows that the minimum in Eq.~(\ref{eq:lsq}) can also be written as
\begin{equation}
\min_{\MAT X \in \FLD C^{N\times N'}}\, \zeta(\MAT X) ~\equiv~ \min_{\MAT X \in \FLD C^{N\times N'}}\, \big\|\, \MAT X\MAT X^\dag \,-\, \MAT M\, \|^2 ~.
\end{equation}
It is easily seen that the gradient matrix of $\zeta(\MAT X)$ is
\begin{equation}
\frac{\FUN d\zeta}{\FUN d\MAT X} ~=~ \frac{\FUN d}{\FUN d\MAT X}\, \TR\Big( {(\MAT X\MAT X^\dag \,-\, \MAT M)}^2 \Big) ~=~ 2\, \big( \MAT X\MAT X^\dag \,-\, \MAT M \big)\, \MAT X ~.
\end{equation}
On setting this to the zero matrix, we obtain the nonlinear matrix equation
\begin{equation}
\label{eq:nme}
\MAT M\, \MAT X ~=~ \MAT X \big( \MAT X^\dag \MAT X \big) ~,
\end{equation}
wherein $\MAT X^\dag\MAT X$ is an $N'\times N'$ Hermitian matrix which, for $N' = 3$ and $\MAT X \in \FLD R^{N\times 3}$, is essentially the inertial tensor (plus a multiple of the identity) of a system of unit mass points with coordinates $\MAT e_i^\top\, \MAT X$ ($0 \le i < N$). Since the Hilbert-Schmidt norm is unitarily invariant, we may assume that these ``coordinates'' have been chosen so that $\MAT X^\dag\MAT X = \DMAT( \xi_1, \ldots, \xi_{N'} )$ is diagonal, in which case Eq.~(\ref{eq:nme}) becomes
\begin{equation}
\label{eq:eve}
\MAT M\, \MAT x_j ~=~ \xi_j\, \MAT x_j \qquad\text{($j = 0, \ldots, N-1$),}
\end{equation}
where $\MAT x_j \equiv \MAT X \MAT e_j$ are the columns of $\MAT X$. It follows that the $\MAT x_j$ are proportional to the eigenvectors $\MAT u_j$ associated with certain nonnegative eigenvalues $\lambda_j = \xi_j$ of $\MAT M$ where, since $\| \MAT x_j \|^2 = \xi_j\,$, the constant of proportionality is $\sqrt{\lambda_j}\,$. On expanding the trace in the function $\zeta(\MAT X)$, we now obtain
\begin{equation} \begin{split}
\zeta(\MAT X) ~=\M[0.5] & \TR\big(\, \MAT M^2 \,-\, 2\, \MAT X\MAT X^\dag\, \MAT M \,+\, (\MAT X\MAT X^\dag)^2 \,\big) \\
=\M[0.5] & \TR\big(\, \MAT M^2 \,\big) \,-\, \TR\big(\, 2\, \MAT X^\dag \MAT M\, \MAT X \,-\, (\MAT X^\dag \MAT X)^2 \,\big) ~.
\end{split} \end{equation}
By Eq.~(\ref{eq:nme}), however, the matrix $\MAT X' \in \FLD C^{N\times N'}$ that minimizes $\zeta(\MAT X)$ satisfies
\begin{equation}
(\MAT X')^\dag\, \MAT M\, \MAT X' ~=~ \big(\, (\MAT X')^\dag \MAT X' \,\big)^2 ~=~ \DMAT\big(\, \lambda_0^2, \ldots, \lambda_{N'-1}^2 \,\big) ~,
\end{equation}
so that
\begin{equation}
\zeta(\MAT X) ~=~ \TR\big(\, \MAT M^2 \,\big) ~-~ \sum_{j=0}^{N'-1}\, \lambda_j^2 ~.
\end{equation}
From this we see that, for any integer $N''$ with $0 \le N'' \le N'$ and $\lambda_j \ge0$ for $0 \le j < N''$, the minimizing $\MAT X'' \in \FLD C^{N\times N''}$ is obtained by setting $\MAT X'' \equiv \big[ \sqrt{\lambda_j}\, \MAT u_j \big]_{j=0}^{N''-1}$. It follows that the minimizing positive semidefinite matrix $\MAT X^\star\,(\MAT X^\star)^\dag$ is obtained by setting $N''$ to the number $N^\star$ of positive eigenvalues of $\MAT M$. QED

\smallskip
This theorem can be used to ``filter'' statistical estimates of either superoperators or their generators so as to obtain a completely positive estimate. In the case of an estimate $\EMB{\ALG S}'$ of a matrix representing an unknown superoperator $\ALG S$, one simply sets any negative eigenvalues of the associated Choi matrix $\EMB{\ALG T \M[0.05]}' = \EMB{Choi}(\EMB{\ALG S}')$ to zero, reconstructs the improved estimate $\EMB{\ALG T\M[0.05] }^\star$ from these modified eigenvalues and the original eigenvectors as in the theorem, and converts the result back into a new estimate $\EMB{\ALG S}^\star = \EMB{Choi}(\EMB{\ALG T}^\star)$ of the superoperator via the same involutory mapping $\EMB{Choi}$. The theorem assures us that this procedure makes the smallest possible change in $\EMB{\ALG T \M[0.05]}'$, with respect to the Hilbet-Schmidt norm, so as to render it positive semidefinite and so ensure that $\EMB{\ALG S}^\star$ represents a completely positive superoperator. Because the mapping $\EMB{Choi}$ simply permutes the elements of its argument, we can be sure that this procedure also minimizes the change $\| \EMB{\ALG S}' - \EMB{\ALG S}^\star \|$ in $\EMB{\ALG S}'$. We now show that $\EMB{\ALG S}^\star$ is assured of being an improved estimate of the corresponding matrix of the true superoperator $\EMB{\ALG S}$, again in the least-squares sense.
\begin{COROLLARY}
For $\EMB{\ALG S}$, $\EMB{\ALG S}'$ and $\EMB{\ALG S}^\star$ defined as above, we have
\begin{equation}
\big\|\, \EMB{\ALG S}^\star \,-\, \EMB{\ALG S} \,\big\| ~\le~ \big\|\, \EMB{\ALG S}' \,-\, \EMB{\ALG S} \,\big\| ~.
\end{equation}
\end{COROLLARY}
\PROOF Since $\EMB{\ALG S}^\star$ is the orthogonal projection of $\EMB{\ALG S}'$ onto the convex cone $\EMB{Choi}(\FLD P)$ of matrices representing completely positive superoperators, $\EMB{\ALG S}' - \EMB{\ALG S}^\star$ is orthogonal to a supporting hyperplane at $\EMB{\ALG S}^\star$, while by its definition $\EMB{\ALG S} \in \EMB{Choi}(\FLD P)$ must be on the opposite side of this hyperplane from $\EMB{\ALG S}'$. This in turn implies that the angle $\theta$ between $\EMB{\ALG S}$ and $\EMB{\ALG S}'$ at $\EMB{\ALG S}^\star$ satisfies $\theta \ge \pi/2$, and hence by the law of cosines
\begin{equation}
0 ~\ge~ \cos(\theta) ~=~ \HALF \big( \| \EMB{\ALG S}^\star - \EMB{\ALG S}' \|^2 \,+\, \| \EMB{\ALG S}^\star - \EMB{\ALG S} \|^2 \,-\, \| \EMB{\ALG S}' - \EMB{\ALG S} \|^2 \big) ~,
\end{equation}
i.e.~$\| \EMB{\ALG S}' - \EMB{\ALG S} \|^2 \ge \| \EMB{\ALG S}^\star - \EMB{\ALG S}' \|^2 + \| \EMB{\ALG S}^\star - \EMB{\ALG S} \|^2 \ge \| \EMB{\ALG S}^\star - \EMB{\ALG S} \|^2$. QED

The procedure in the case of a QDS generator $\ALG F = \ALG G + \imath\ALG H$ is a bit more involved, since one needs to compute the projection of the Choi matrix $\EMB{\ALG E}' \equiv \EMB{\ALG P}^{\MAT I}\, \EMB{Choi}( \EMB{\ALG F}' )\, \EMB{\ALG P}^{\MAT I}$ of the estimate $\EMB{\ALG F}'$ before diagonalizing it. This of course will remove the Hamiltonian superoperator component, which must then be obtained by some other means. In addition, one cannot reconstruct a matrix $\EMB{\ALG G}^\star$ for the decoherent component $\ALG G$ of $\ALG F$ from the matrix $\EMB{\ALG E}^\star$ obtained by setting any negative eigenvalues $\varepsilon_m$ of $\EMB{\ALG E}'$ to zero simply by applying the $\EMB{Choi}$ mapping, since the other terms needed to preserve the trace will also have been lost in the projection (if indeed the estimate $\EMB{\ALG F}'$ itself were trace-preserving). Instead, one has to construct all the Lindblad operators $\MAT L_m$ such that $\COL(\MAT L_m) = \sqrt{\varepsilon_m} \EMB{v}_m$, where $\varepsilon_m > 0$, $\EMB{v}_m$ are eigenvalue, eigenvector pairs of $\EMB{\ALG E}'$, and compute $\EMB{\ALG G}^\star$ as indicated in Eq.~(\ref{eq:lindgen}). As a result, there is no guarantee that $\EMB{\ALG G}^\star$ will be closer to its true value $\EMB{\ALG G}$ versus the Hilbert-Schmidt norm, although we expect that this will usually be the case. Further discussion regarding how one might go about solving these problems must take the exact experimental situation at hand into account, and as such is outside the scope of this paper.

In the remainder of this section we will illustrate how the above results may be applied to a simple example, namely the \emph{Bloch equations} for a single spin $1/2$ qubit in a frame rotating at its Larmour frequency in an applied magnetic field \cite{ErnBodWok:87}. As is well-known \cite{ViFoLlTsCo:00}, these can be expressed in canonical Lindblad form as
\begin{align}
\label{eq:bloch}
\dot\RHO ~=~ \ALG L(\RHO) ~\equiv~ \notag
& \tfrac{1+\Delta}{4T_1}\,  \big( 2\, \MAT E_{01}\,\RHO\,\MAT E_{10} - \MAT E_{00}\,\RHO - \RHO\,\MAT E_{00} \big) ~+ \\
& \tfrac{1-\Delta}{4T_1}\, \big( 2\, \MAT E_{10}\,\RHO\,\MAT E_{01} - \MAT E_{11}\,\RHO - \RHO\,\MAT E_{11} \big) ~+ \\ \notag
& \Big( \tfrac1{2T_2} - \tfrac1{4T_1} \Big) \big( (\MAT E_{00} - \MAT E_{11})\, \RHO\, (\MAT E_{00} - \MAT E_{11}) - \RHO \big) ~,
\end{align}
where $T_1$ and $T_2$ are the characteristic relaxation and decoherence times and $\Delta = p_0 - p_1$ is the excess probability in the ground state $\MAT E_{00}$ at equilibrium. The supermatrix of the generator versus a  Hilbert space basis in the ordering $\MAT E_{00\,}, \MAT E_{10\,}, \MAT E_{01\,}, \MAT E_{11}$ induced by the ``$\COL$'' operator is
\begin{align}
\label{eq:gen}
& \tfrac{1+\Delta}{4T_1}\, \big( 2\, \MAT E_{01} \otimes \MAT E_{01} - \MAT I \otimes \MAT E_{00} - \MAT E_{00} \otimes \MAT I \big) ~+ \notag \\
& \tfrac{1-\Delta}{4T_1}\, \big( 2\, \MAT E_{10} \otimes \MAT E_{10} - \MAT I \otimes \MAT E_{11} - \MAT E_{11} \otimes \MAT I \big) ~+ \notag \\
& \Big( \tfrac1{2T_2} - \tfrac1{4T_1} \Big) \Big( \big( \MAT E_{00} - \MAT E_{11} \big) \otimes \big( \MAT E_{00} - \MAT E_{11} \big) - \MAT I \otimes \MAT I \Big) \\ =~ \notag
& \begin{bmatrix}
-\tfrac{1-\Delta}{2T_1} & 0 & 0 & \tfrac{1+\Delta}{2T_1} \\
0 & -\tfrac1{T_2} & 0 & 0\\ 0 & 0 & -\tfrac1{T_2} & 0\\
\tfrac{1-\Delta}{2T_1} & 0 & 0 & -\tfrac{1+\Delta}{2T_1}
\end{bmatrix} \M\equiv\M \EMB{\ALG L} ~.
\end{align}
The time-dependent exponential of this matrix may be shown to be
\begin{equation}
\label{eq:expm}
\EMB{Exp}\big(- \EMB{\ALG L}\,t \big) ~=~ \HALF
\left[ \begin{smallmatrix}
\big( 1 + e^{-t/T_1} \big) + \Delta \big( 1 - e^{-t/T_1} \big)
& 0 & 0 &
\big( 1 - e^{-t/T_1} \big) + \Delta \big( 1 - e^{-t/T_1} \big)
\\ 0 & 2\,e^{-t/T_2} & 0 & 0\\ 0 & 0 & 2\,e^{-t/T_2} & 0\\
\big( 1 - e^{-t/T_1} \big) - \Delta \big( 1 - e^{-t/T_1} \big)
& 0 & 0 &
\big( 1 + e^{-t/T_1} \big) - \Delta \big( 1 - e^{-t/T_1} \big)
\end{smallmatrix} \right] ,
\end{equation}
which in turn corresponds to the Choi matrix
\begin{equation}
\EMB{\ALG M}(t) ~\equiv~ \HALF
\left[ \begin{smallmatrix} \X[2]
( 1 + e^{-t/T_1} ) + \Delta\, ( 1 - e^{-t/T_1} )\!
& 0 & 0 & 2\,e^{-t/T_2} \\ 0 &
( 1 - e^{-t/T_1} ) (1 - \Delta)
& 0 & 0\\ 0 & 0 &
( 1 - e^{-t/T_1} ) (1 + \Delta)
& 0\\ 2\,e^{-t/T_2} & 0 & 0 &
\!( 1 + e^{-t/T_1} ) - \Delta\, ( 1 - e^{-t/T_1} )\\
\end{smallmatrix} \right] .
\end{equation}
This in turn is readily shown to be positive semidefinite for all $t \ge0$ if $2T_1 \ge T_{2\,}$. Its derivative at $t = 0$, however, is
\begin{equation}
\dot{\EMB{\ALG M}\,}\!(0) ~=~
-\EMB{Choi}(\EMB{\ALG L}) ~=~
\HALF \begin{bmatrix}
-\tfrac{1-\Delta}{2T_1} & 0 & 0& -\tfrac1{T_2}\\
0 & \tfrac{1-\Delta}{2T_1} & 0 & 0\\ 0 & 0 & \tfrac{1+\Delta}{2T_1}& 0\\
-\tfrac1{T_2} & 0 & 0 & -\tfrac{1+\Delta}{2T_1}
\end{bmatrix} ~,
\end{equation}
and the outermost $2 \times 2$ block of this matrix is positive semidefinite \emph{only} if $2T_1 \le T_{2\,}$. Applying the projection $\EMB{\ALG P}^{\MAT I} \,\equiv\, \EMB{\ALG I} - \COL(\MAT I)\,\COL^\dag(\MAT I)/2$ converts it to
\begin{equation}
-\EMB{\ALG P}^{\MAT I}\,\EMB{Choi}(\EMB{\ALG L})\,\EMB{\ALG P}^{\MAT I} ~=~
\begin{bmatrix} \tfrac1{2T_2}-\tfrac1{4T_1} & 0 & 0 & \tfrac1{4T_1}-\tfrac1{2T_2} \\
0 & ~\tfrac{1-\Delta}{2T_1}~ & 0 & 0\\ 0 & 0 & ~\tfrac{1+\Delta}{2T_1}~ & 0\\
\tfrac1{4T_1}-\tfrac1{2T_2} & 0 & 0 & \tfrac1{2T_2}-\tfrac1{4T_1} \end{bmatrix} ~,
\end{equation}
which is now positive semidefinite with eigenvalue, eigenvector pairs:
\begin{equation}
\bigg( 0 ,~ \left[ \begin{smallmatrix} 1\\0\\0\\1 \end{smallmatrix} \right] \bigg),\quad \bigg( \tfrac1{2T_2} - \tfrac1{4T_1} ,~ \left[ \begin{smallmatrix} ~1\\~0\\~0\\-1 \end{smallmatrix} \right] \bigg),\quad \bigg( \tfrac{1+\Delta}{2T_1} ,~ \left[ \begin{smallmatrix} 0\\1\\0\\0 \end{smallmatrix} \right] \bigg),\quad \bigg( \tfrac{1-\Delta}{2T_1} ,~ \left[ \begin{smallmatrix} 0\\0\\1\\0 \end{smallmatrix} \right] \bigg) .
\end{equation}
The eigenvectors are easily seen to be obtained by applying the ``$\COL$'' operator to the matrices $\MAT I = \MAT E_{00} + \MAT E_{11\,}$, $\MAT E_{00} - \MAT E_{11\,}$, $\MAT E_{10}$ and $\MAT E_{01\,}$, returning us to the canonical Lindblad form in Eq.~(\ref{eq:bloch}).

We will now use this example to illustrate how the matrix formulae obtained in this paper can be applied to QPT, by numerically simulating the ``data'' needed for QPT from the above solution to the Bloch equations. These data correspond to an experimental scenario in which a set of precisely known input states ${\{ \RHO_k^\LAB{in} \}}_{k=1}^K$ were allowed to evolve under the propagator in Eq.~(\ref{eq:expm}) for varying periods of time, and the results ${\{ \RHO_k^\LAB{out} \}}_{k=1}^K$ determined by \emph{state} tomography \cite{Leonhardt:97, ChiChuLeu:01, DAriaPrest:01, NielsenChuang:01, HCLBFPTWBH:02}. Assuming that the input states span the space of single-qubit Hermitian operators, this allows us to determine the propagators at each time point according to
\begin{equation} \begin{split}
& \EMB{Exp}\big( -\EMB{\ALG L}\, t \big) \big[ \COL(\RHO_1^\LAB{in}), \ldots, \COL(\RHO_K^\LAB{in}) \big] ~=~ \big[ \COL(\RHO_1^\LAB{out}), \ldots, \COL(\RHO_K^\LAB{out}) \big] \\
\Leftrightarrow~ & \EMB{Exp}\big( -\EMB{\ALG L}\, t \big) ~=~ \big[ \COL(\RHO_1^\LAB{out}), \ldots, \COL(\RHO_K^\LAB{out}) \big] \big[ \COL(\RHO_1^\LAB{in}), \ldots, \COL(\RHO_K^\LAB{in}) \big]^{-1} ~.
\end{split} \end{equation}
Although this relation is exact when the output states are known precisely, in actual practice experimental errors would result in only an approximate estimate $\EMB{\ALG S}'(t)$ of the actual propagator $\EMB{\ALG S}(t) \equiv \EMB{Exp}( -\EMB{\ALG L}\, t)$. If one obtains such estimates at an arithmetic sequence of time points $0 = t_0, t_1, \ldots, t_J = J t_1$, however, one may solve a linear least-squares problem to obtain an improved estimate of the propagator $\EMB{\ALG S}_1 \equiv \EMB{\ALG S}(t_1)$ at the first nonzero time point \cite{NaDaWaHa:97}, namely
\begin{equation}
{\min}_{\EMB{\ALG T}}\, \big(\chi(\EMB{\ALG T})\big) \quad\text{where}\quad \chi(\EMB{\ALG T}) ~\equiv~ \sum_{j\,=\,0}^{J-1}\, \big\| \EMB{\ALG T}\, \EMB{\ALG S}_j' \,-\, \EMB{\ALG S}_{j+1}' \,\big\|^2 ~.
\end{equation}
One may of course set $\EMB{\ALG S}_0' = \EMB{\ALG S}(t_0) = \MAT I \otimes \MAT I$, the $4\times4$ identity, and one should also filter the remaining estimates by symmetrizing their Choi matrices (i.e.~by adding them to their adjoints and dividing by two), setting any negative eigenvalues $\psi = 0$ and transforming back to a new estimate (as described previously). The minimizing solution to this least-squares problem is easily shown to be
\begin{equation}
\EMB{\ALG S}_1'' ~\equiv~ \Big( \sum_{j\,=\,1}^{J-1}\, \EMB{\ALG S}_j'\, {(\EMB{\ALG S}_{j-1}')}^\dag \Big) \Big( \sum_{j\,=\,1}^{J-1}\, \EMB{\ALG S}_j'\, {(\EMB{\ALG S}_j')}^\dag \Big)^\ddag ~,
\end{equation}
where in most cases the Moore-Penrose inverse ($\ddag$) may be replaced by the usual matrix inverse \cite{LawsonHanson:95}.

Finally, $\EMB{\ALG S}_1''$ may be converted into an estimate of the generator via the matrix ``pseudo-logarithm'', $\EMB{Plog}$. This is computed by diagonalizing $\EMB{\ALG S}_1'' = \EMB{\ALG W}\, \EMB{\Phi}\, \EMB{\ALG W}^{-1}$, setting any eigenvalues $\phi_i \le 0$ or $\phi_i \ge 1$ to zero while taking the usual logarithm of the rest, then performing the inverse similarity transformation and dividing by $t_1$, i.e.
\begin{equation}
t_1\, \EMB{\ALG L}'' ~=~ \EMB{Plog}\big(\EMB{\ALG S}_1''\big) ~\equiv~ \EMB{\ALG W}\, \EMB{Plog}\big(\EMB{\Phi}\big)\, \EMB{\ALG W}^{-1} ~=~ \sum_{i\,=\,0}^{3}\, \FUN{plog} (\phi_i)\, \big(\EMB{\ALG W}\, \MAT e_i\big) \big(\EMB{\ALG W}^{-1} \MAT e_i\big)^\dag ~,
\end{equation}
where
\begin{equation}
\FUN{plog}(\phi_i) ~\equiv~ \begin{cases} \log(\phi_i) & \text{if}~0 < \phi_i < 1; \\ 0 & \text{otherwise.} \end{cases}
\end{equation}
The eigenvalues will be real since no Hamiltonian was assumed in the simulations, and
arguments similar to those involved in Theorem \ref{thm:eyh} can be used to show that the pseudo-logarithm will then yield a generator $\EMB{\ALG L}''$ that minimizes $\| \EMB{\ALG S}_1'' - \EMB{Exp}(-\EMB{\ALG L}''t_1) \|$. Lastly, the estimate $\EMB{\ALG L}''$ is filtered by projecting its symmetrized Choi matrix by $\EMB{\ALG P}^{\MAT I}$, setting any eigenvalues $\varepsilon = 0$, and reconstructing to obtain the optimum estimate $\EMB{\ALG L}^\star$, as described above.

The specific values of the parameters used for the simulations were $T_1 = 0.5$, $T_2 = 0.1$ and $\Delta = 0.1$; the relaxation times $T_1$ and $T_2$ are typical of liquid-state NMR samples, while the polarization $\Delta$ was deliberately made larger to render it visible despite the noise.
In accord with Eq.~(\ref{eq:gen}), these gave rise to the generator
\begin{equation}
\EMB{\ALG L} ~\equiv~
\left[ \begin{smallmatrix} -0.9 & 0 & 0 & 1.1 \\[3pt] 0 & -10.0 & 0 & 0 \\[3pt]
0 & 0 & -10.0 & 0 \\[3pt] 0.9 & 0 & 0 & -1.1 \end{smallmatrix} \right]
\end{equation}
The input states were taken to be $\MAT E_{00}$, $\MAT E_{11}$, $(\MAT e_0 + \MAT e_1) (\MAT e_0 + \MAT e_1) / 2$, and $(\MAT e_0 - \imath\MAT e_1) (\MAT e_0 + \imath\MAT e_1) / 2$, while the times used were set to $t_j \equiv j / 4$ ($j = 0,\ldots,4$). Finally, the noise levels evaluated were $\Omega_1 = 0.01$, $\Omega_2 = 0.05$ and $\Omega_3 = 0.25$, where the noise was simply added to the output states $\{ \RHO^\LAB{out}_k \}$ with a Gaussian distribution, zero mean, and variances $\sigma_j^2\,\Omega_k^2$ proportional to the mean-square size $\sigma_j^2$ of the elements of $\EMB{\ALG S}(t_j)$ ($j = 0,\ldots,J; k = 1,2,3$). The results below were averaged over $100$ independent estimations of the propagators at each time point, using different random noise for each estimation and time point, followed by filtering and fitting to obtain estimates of the generator, all at each of the three specified noise levels.

\begin{table}
{\setlength{\fboxrule}{3pt} \setlength{\fboxsep}{-3pt} \fbox{
\begin{tabular}{c||cccc|c}
$\qquad$ & $\qquad t_1\qquad$ & $\qquad t_2\qquad$ & $\qquad t_3\qquad$ &
$\qquad t_4\qquad$ & $~|\{ \psi < 0 \}|~$ \\ \hline\hline
$~\Omega_1~$ & 0.0108 & 0.0121 & 0.0116 & 0.0127 & 0.000 \\
$~\Omega_2~$ & 0.0581 & 0.0601 & 0.0644 & 0.0605 & 0.000 \\
$~\Omega_3~$ & 0.3062 & 0.3038 & 0.3074 & 0.3098 & 0.290 \\
\end{tabular}
}}
\caption{Average over 100 runs of Hilbert-Schmidt norms of the changes in the propagators on symmetrizing and filtering the eigenvalues $\{ \psi \}$ of their Choi matrices, divided by the norm of the actual propagator $\|\EMB{\ALG S}\|$ (see text); the last column shows the average number $|\{ \psi < 0 \}|$ of negative eigenvalues of that were set to zero.}
\end{table}

Table 1 shows the average changes made to the propagator estimates upon symmetrizing and filtering the eigenvalues of their Choi matrices, as measured by the Hilbert-Schmidt norm of the difference divided by that of the true propagator, together with the average number $|\{ \psi < 0 \}|$ of eigenvalues set to zero in the process. It may be seen that the changes in the estimated propagators upon filtering became significant as the noise level increased, but were generally little more than the added noise. Negative eigenvalues were frequently encountered only at the highest noise level $\Omega_3 = 0.25$, however, so in fact most of these changes were due to the symmetrization needed to make the estimated Choi matrices Hermitian.

Table 2 shows the average changes made to the various generator estimates computed (this time normalized by the norm of the true generator), together with the average numbers of eigenvalues set to zero in computing the pseudo-logarithm ($|\{ \phi < 0 \}|$) and in filtering ($|\{ \varepsilon < 0 \}|$). Again, few eigenvalues with incorrect signs were encountered either in computing the pseudo-logarithm, or in symmetrizing and filtering the resulting generators. This means that, once again, most of the improvement was obtained via the projection $\EMB{\ALG P}^{\MAT I}$ and subsequent reconstruction, forcing the estimated generators $\EMB{\ALG L}^\star$ to preserve the trace (which the unfiltered estimates $\EMB{\ALG L}''$ did not). Finally, it should be noted that the filtered generators $\EMB{\ALG  L}^\star$ usually came out closer to the actual solution than the unfiltered, although this was not invariably so. Together, these numerical results strongly support our claim that the formulae derived in this paper provide a powerful set of tools with which to tackle quantum process tomography on systems that may be aptly modeled as a quantum dynamical semigroup.

\begin{table}
{\setlength{\fboxrule}{3pt} \setlength{\fboxsep}{-3pt} \fbox{
\begin{tabular}{c||ccc|cc}
$\qquad$ & $~\| \EMB{\ALG L}'' - \EMB{\ALG L}^\star \|~$ & $~\| \EMB{\ALG L}'' - \EMB{\ALG L} \|~$ & $~\| \EMB{\ALG L}^\star - \EMB{\ALG L} \|~$  & $~|\{ \phi < 0 \}|~$ & $~|\{ \varepsilon < 0 \}|~$ \\ \hline\hline
$~\Omega_1~$ & 0.0077 & 0.0305 & 0.0300 & 0.000 & 0.000 \\
$~\Omega_2~$ & 0.0634 & 0.1720 & 0.1676 & 0.010 & 0.420 \\
$~\Omega_3~$ & 0.2971 & 0.6355 & 0.5553 & 0.580 & 0.840 \\
\end{tabular}
}}
\caption{Average Hilbert-Schmidt distances (columns 1--3) among the estimates of the generators divided by the norm of the actual generator $\|\EMB{\ALG L}\|$, and (columns 4 -- 5) average numbers of eigenvalues set to zero in obtaining these estimates (see text).}
\end{table}

\bigskip
\section{Conclusions}
In this paper we have presented formulae by which the supergenerators and superpropagators of quantum dynamical semigroups may be manipulated, placed in canonical Lindblad and Kraus form, and all these forms interconverted. These formulae constitute a set of tools that should be particularly valuable in developing robust procedures for quantum process tomography \cite{NielsenChuang:01} and quantum channel identification \cite{Fujiwara:01}, using diverse forms of experimental data. We have illustrated one such application using data simulated from the well-known Bloch relaxation equations on a single spin $1/2$ qubit \cite{ErnBodWok:87}, which assumed that full state tomography versus a basis of input states could be performed. This example demonstrated the anticipated robustness of the procedures employed, which was the result of combining the eigenvalue characterizations of completely positive supergenerators and superpropagators derived in this paper with powerful matrix approximation methods derived from the field of principal component analysis \cite{Joliffe:86}.

It should be clearly understood, nevertheless, that the procedures given here were intended primarily to provide a concrete example of how the mathematical results given in this paper can be applied to quantum process tomography, and not as a prescriptive recipe that is in all cases optimal --- or even applicable. For example, the system of interest will often evolve coherently as it relaxes towards equilibrium, and at a rate far larger than the relaxation processes themselves. In this case the relaxation generator itself will be averaged, significantly complicating its physical interpretation, and the superpropagators determined from full state tomography versus an input basis set will usually have complex eigenvalues. Even assuming its matrix can be fully diagonalized, the well-known ambiguity of the matrix logarithm with respect to the addition on arbitrary multiples of $2\pi\imath$ onto its eigenvalues will render our ``pseudo-logarithm'' technique inapplicable. Particularly in such cases, better results can be expected from nonlinear fits of the supergenerator to the superpropagators \cite{NajfeHavel:95a, NaDaWaHa:97}, but the question of whether these problems are best solved by computational means, experimental means, or some combination thereof, will clearly depend upon the circumstances.

There are further many other ways to represent a quantum state besides a density matrix, for example by a Wigner distribution \cite{Leonhardt:97}, or it may even be desirable to forgo state tomography altogether and to base quantum process tomography on a sequence of time-dependent observations which, although individually insufficient to fully determine the superoperator or even the system's quantum state, nevertheless do so in aggregate. Alternatively, one might utilize a form of indirect measurement via qubits outside of, but interacting with, the system of interest \cite{ErnBodWok:87, Fujiwara:01, Leung:02}. We anticipate that many creative applications and extensions of the techniques introduced in this paper will be developed in the years ahead, as quantum information processing technologies progress towards experimental reality.

\pagebreak[4]
{\setlength{\parindent}{0pt}
\begin{acknowledgments}
\vspace{-6pt} \noindent
This work was supported by ARO grants DAAD19-01-1-0678 and DAAD19-01-1-0519, and by DARPA grant MDA972-01-1-0003. The author thanks David G. Cory, Nicolas Boulant and Lorenza Viola for helpful conversations, Mary-Beth Ruskai for pointing out the connection of this work to that of Choi \cite{Choi:75}, and an anonymous referee for suggesting structural improvements to the manuscript. After this paper was submitted, a preprint by D.~W.\linebreak[2]~Leung \cite{Leung:02} came to our attention, which also proposes using some of the results in section \ref{sec:CKOSR} for quantum process tomography.
\end{acknowledgments}
}\pagebreak[2]

\bibliography{phys,math,csci,self,nmr}

\end{document}
\end

%% file: macros.tex
\makeatletter
\def\@tempa{revtex4}
\newif\ifisRevTeX
\ifx\class@name\@tempa
   \isRevTeXtrue
\else
   \isRevTeXfalse
\fi
\makeatother

\newcommand{\EMB}{\boldsymbol}

\newcommand{\LAB}[1]{{\mathsf{#1}}}
\newcommand{\FUN}[1]{{\mathrm{#1}}}
\newcommand{\MAT}[1]{{\mathbf{#1}}}

\newcommand{\FLD}[1]{{\mathbb{#1}}}
\newcommand{\ALG}[1]{{\mathcal{#1}}}

\newcommand{\OL}{\overline}

\newcommand{\HALF}{\mathchoice{{\textstyle\frac12}}{\frac12}{\frac12}{\frac12}}

\newcommand{\M}[1][1]{\hspace{#1em}}
\newcommand{\DAB}[1]{\rule[0pt]{0pt}{#1}}
\newcommand{\X}[1][1]{\DAB{#1ex}}

\newcommand{\COL}{\MAT{col}}
\newcommand{\DMAT}{\MAT{Diag}}

\newcommand{\TR}{\FUN{tr}}

\newcommand{\RHO}{{\EMB\rho}}

\newtheorem{THEOREM}{Theorem}
\newtheorem{LEMMA}[THEOREM]{Lemma}
\newtheorem{COROLLARY}[THEOREM]{Corollary}
\newtheorem{PROPOSITION}[THEOREM]{Proposition}
\newcommand{\PROOF}[1][{}]{\subparagraph{\ifisRevTeX{\hspace{-\parindent}}\fi\textbf{Proof#1.}}}